\documentclass{aastex}          
\usepackage{spr-astr-addons}    
\usepackage{url}

\RequirePackage{color}
\shorttitle{Search of variable stars in the northern sky}
\shortauthors{Pak\v stien\.e et al.}

\begin{document}


\title{Search for variable stars in the northern sky:\\Analysis of photometric time series for 3598 stars
}


\author{E. Pak\v{s}tien\.{e}\altaffilmark{1}} \email{erika.pakstiene@tfai.vu.lt}  \author{R. Janulis\altaffilmark{}} \author{A. Drazdauskas\altaffilmark{}}  
\author{\v S. Mikolaitis\altaffilmark{}} \author{G. Tautvai\v{s}ien\.{e}\altaffilmark{}} \author{R. Minkevi\v{c}i\={u}t\.{e}\altaffilmark{}} \author{V. Bagdonas\altaffilmark{}}
\affil{Astronomical Observatory, Institute of Theoretical Physics and Astronomy, Vilnius University, Saul\.{e}tekio av. 3, 10257 Vilnius, Lithuania, {erika.pakstiene@tfai.vu.lt}}

\author{L. Klebonas\altaffilmark{}}
\affil{Astronomical Observatory, Institute of Theoretical Physics and Astronomy, Vilnius University, Saul\.{e}tekio av. 3, 10257 Vilnius, Lithuania \\ Mathematisch-Naturwissenschaftliche Fakult\"at, Universit\"at Bonn, Wegelerstra{\ss}e 10, 53115 Bonn, Germany}
\author{J.K.T. Qvam\altaffilmark{}}
\affil{Horten Videreg\aa ende Skole, Bekkegata 2, 3181 Horten, Norway}
\altaffiltext{1}{e-mail of the corresponding author: erika.pakstiene@tfai.vu.lt}



\begin{abstract}
As extensive ground-based observations and characterisation of different variable stars are of the utmost importance in preparing optimal input catalogues for space missions, 
our aim was to search for new variable stars in selected fields of the northern sky. 
We obtained 24\,470 CCD images and analysed photometric time series of stars using the DAOPHOT based package Muniwin as the first step, and the Period04 package was used to further analyse the suspected new variable stars. 
The light curves and other observational results are presented for 3598 stars online. 
We found 81 new variable stars, among them is an 
eclipsing binary with a variable component and possibly eccentric orbits TYC4038-693-1 which we also observed spectroscopically, four $\delta$~Scuti candidates, six other variable stars with periods falling into the interval of 35 minutes to 20 days. Furthermore, we identified 70 slowly varying stars with so far undefined periodicity. Additional photometric and spectral observations were  carried out for  TYC\,2764-1997-1, and its previous candidacy for eclipsing binaries was approved. 
\end{abstract}





\section{Introduction}

This work is part of our Spectroscopic and photometric survey of the northern sky (SPFOT;  \citealt{Mikolaitis2018}; \citealt{Pakstiene2018_deltaScuti}) started at the Mol\.{e}tai Astronomical Observatory (MAO) in 2016. It aims to deliver a full spectroscopic characterisation and variability information of the brightest stars in the northern sky. Bright stars will soon be targeted by the NASA TESS and later on by the ESA PLATO space missions. 

\addtolength{\tabcolsep}{17.2pt}
\begin{table*}
\caption{Information on the observed fields}              
\centering                          
\begin{tabular}{l c c c c}        
\hline\hline                 
Central star     &  $\alpha$(2000)   & $\delta$(2000)     & Preliminary  	& FOV 	\\
of the field    &      [h m s]          &   [$^\circ$  $^\prime$  $^{\prime\prime}$]  & PLATO field & [deg$^2$]  \\
\hline 
HIP 2923   & 00 37 03.56 &+31 29 11.31  & STEP07  	&0.39	  		\\
HIP 5526   & 01 10 43.31 &+27 52 04.61  &STEP07	    &0.39	  	\\
HIP 5659   & 01 12 41.26 &+65 00 32.83  & STEP02	&1.17  	  	\\
HIP 11090  & 02 22 50.30 &+41 23 46.67  &STEP07	    &0.39	  \\
HIP 17585  & 03 46 00.94 &+67 12 05.78  &  STEP02	&0.39	  	\\
HIP 74155  & 15 09 06.24 &+69 39 11.11 &    STEP02	&0.39	  		   \\
HIP 101473 & 20 33 53.70 &+10 03 35.05 &    STEP05  &0.39	  	\\
HIP 106219 & 21 30 53.28 &+24 46 51.90 &    STEP05	&0.39	 \\
HIP 106223 & 21 30 57.05 &+16 34 15.57  & STEP05	&0.39	 	\\
HIP 107786 & 21 50 08.23 &+19 25 26.38  & STEP05	&0.39	  \\
HIP 113487 & 22 58 58.96 &+34 04 00.67  & STEP05	&0.39	  		  \\
HIP 115093 & 23 18 42.26 &+36 05 24.81  & STEP05	&0.39	  	\\
HIP 115856 & 23 28 23.54 &+19 53 08.09  & STEP05 	&0.39	  	\\  
\hline                                   
\end{tabular}
\end{table*}
\label{Table1}
\addtolength{\tabcolsep}{-18pt}

This paper provides description of the photometric variability analysis results of 3598 stars and presents an online catalogue of their stellar light curves. We aimed at identifying new variable stars and to characterise them if possible.  We also targeted an interesting eclipsing binary candidate TYC\,2764-1997-1 and performed its  photometric and spectrometric observations. 
Complementary observations and analyses were carried out for the eclipsing binaries ASASSN-V J011509.99+652848.0, J213047.63+161528.6, and J215042.53+193829.5 listed in the ASAS-SN Variable Stars Database (https://asassn.osu.edu/variables, \citealt{Jayasinghe2018_II}, \citealt{Jayasinghe2018}). 

\section{Observations}


Observations were performed at the Mol\.{e}tai Astronomical Observatory (MAO, Lithuania) with a 51~cm Maksutov-type telescope with 35~cm working diameter of primary mirror and the Apogee Alta U47 CCD camera. This telescope was chosen for the observations since it has quite a large field of view (37.5 x 37.5~arcmin$^2$ or 0.39~deg$^2$) and allows us to observe bright stars without saturation of CCD pixels. 
For the observations we used the $Y$ filter of a medium-band Vilnius photometric system.
Its effective wavelength is at 466~nm and the width is 26~nm (\citealt{Straizys&Sviderskiene1972}) which is close to the Johnson's $B$ filter, but is transparent for a narrower range of wavelengths.

The observations were carried out in a semi-robotic mode. 
 During the night, we observed light curves of stars in $5-7$ different fields in the sky with a cadence of $15-30$\,min. Every field was observed with several different exposure times (short exposures for the variability analysis of brighter stars and the longer ones in order to analyse fainter objects). We were taking at least three images with each exposure, thus we have collected between 9 and 15 images for each field every $15-30$\,min.
 We were taking more than 10 images of the bias, dark and sky flat fields during each night for the CCD image calibration.

\addtolength{\tabcolsep}{7pt}
\begin{table*}
\caption{Information on the collected data and the new variable candidates found in the fields.}    
\centering                          
\begin{tabular}{l r c c c c c}        
\hline\hline                 
Central star & Date of the session~  &Images  &Runs  &Stars  & Limiting & New    \\
of the field&  JD--2457500 ~ ~ ~ &      &   &   & mag &  variables\tablenotemark{a} \\
\hline 
HIP 2923   &120.35181--149.41487 & 2965	&17	&169	&  15.6 	&1 PV			\\
HIP 5526   &120.37205--144.57982 & 1684	&12	&120	&	15.4	& --			\\
HIP 5659   &120.35375--149.41867 & 2105	&24 &1015	&	15.5	&6 PV, 1 EB \\
           &                     &      &    &      &           & 41 SSVS \\ 
HIP 11090  &125.30497--149.41965 & 1682	&12	&457	&	15.8	& 4 SSVS		\\
HIP 17585  &132.27223--149.42166 & 3475	&14	&139	&	14.9	&1 PV		\\
HIP 74155  &97.40562--120.45030 & 175	&6  &76    	&	14.8	&4 SSVS		   \\
HIP 101473 &113.39664--131.52517 & 675	&7	&361	&	15.0	&3 SSVS	\\
HIP 106219 &97.40923--131.56192 & 1312	&14	&180	&	14.4	&1 PV, 6 SSVS	\\
HIP 106223 &97.40964--149.42793 & 3531	&28	&206	&	15.4	&1 PV, 2 SSVS		\\
HIP 107786 &97.41021--131.55910 & 1215	&14	&233	&	15.1	& --	\\
HIP 113487 &97.41116--131.56308 & 519	&11 &252	&	15.2	&2 SSVS		  \\
HIP 115093 &113.38308--131.56482 & 1390	&10	&271	&	15.4	& 7 SSVS	\\
HIP 115856 &113.38482--149.42956 & 3742	&17	&119	&	15.4	&1 SSVS	\\   
\hline                                   
\multicolumn{6}{l}{$^a$\small EB -- eclipsing binary, PV -- periodic variable, and SSVS -- suspected slowly varying stars}
\centering
\end{tabular}
\label{Table2}
\end{table*}
\addtolength{\tabcolsep}{-7pt}

In Table~\ref{Table1} a list of observed fields is presented, including 
their central stars, the information to which of the preliminary PLATO fields they correspond, and the field-of-view (FOV). The field around HIP\,5659 was large since we found many variable stars on edges of the initial smaller field. In Table~\ref{Table2} we present the number of runs, the number of stars for which the light curves were obtained, and the limiting $Gaia$ $G$ magnitude of investigated stars in the field. The last column shows how many new variable star candidates  were detected and their expected types (EB -- Eclipsing Binary, PV -- Periodic Variable, and SSVS -- Suspected Slowly Varying Star).

In addition, two stars were observed spectroscopically. The eclipsing binary candidate TYC\,2764-1997-1 was observed on the Nordic Optical Telescope (NOT). We performed its quasi-simultaneous photometric  observations  with the Andalucia Faint Object Spectrograph and Camera (ALFOSC) and spectral observations with the FIbre-fed \'{E}chelle Spectrograph (FIES; \citealt{Frandsen1999}; \citealt{Telting2014}).

Quasi-simultaneous photometric and spectroscopic observations were also made for the discovered eclipsing binary star TYC4038-693-1. Vilnius University Echelle Spectrograph VUES (\citealt{Jurgenson2014,Jurgenson2016}) installed on the MAO 1.65~m telescope was used for spectroscopic observations of this star.

\section{Data reduction and analysis}

The observed images were processed and the light curves of all detected stars in the fields were derived with a Muniwin program from the software package C-Munipack\footnote{http://c-munipack.sourceforge.net/}  (\citealt{Muniwin14}), as it was described in \citet{Pakstiene2018_deltaScuti}. The Muniwin program is built on the  basis of software package DAOPHOT for stellar photometry in crowded stellar fields (\citealt{Daophot87}) and is designed for the time series differential aperture photometry and search for variable stars. 

We used the Muniwin program to determine the instrumental magnitudes of all detected stars in the field. Then we calculated amplitude spectra of the stars and selected among them one comparison star per field, which did not show any signal indicating its variability. Names of the used comparison stars are listed and their amplitude spectra are presented in our previous paper on the $\delta$~Scuti candidates in the same fields (\citealt{Pakstiene2018_deltaScuti}). For the further analysis, we calculated differential magnitudes of the targets using the selected comparison stars.

\begin{figure}[t]
    \includegraphics[width=\hsize]{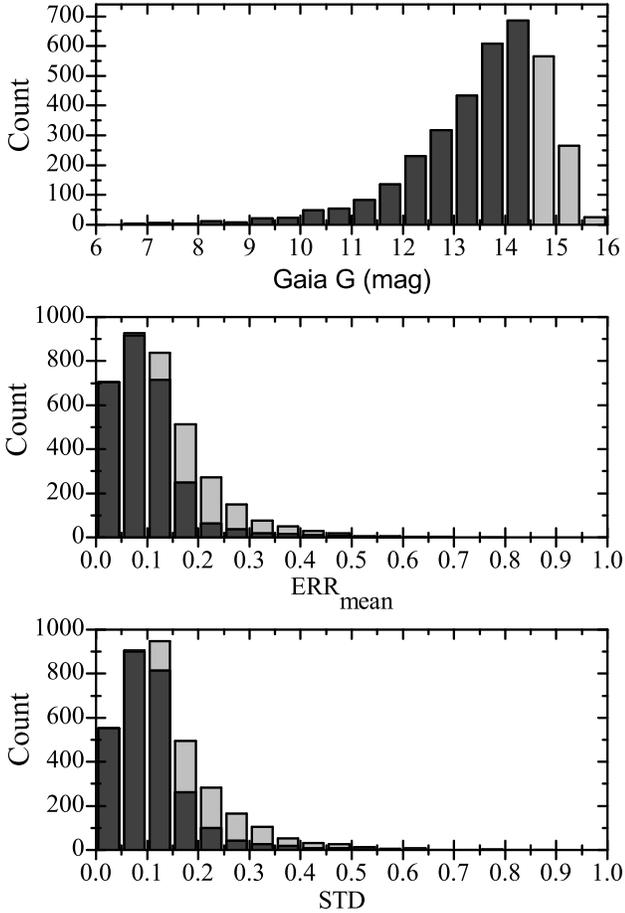}
\caption{Characteristics of the observed set of stars and quality of the obtained light curves.      }
\label{histogram}
   \end{figure}

We analysed a data set of 3598 stars. The upper panel of Fig.~\ref{histogram} shows a distribution of Gaia G magnitudes of the observed stars. The faintest stars are of 15.8~mag in the Vilnius photometric system $Y$ filter, however according to the distribution of magnitudes we may expect reliable results for stars up to 14.5~mag, where the maximum at the histogram appears. The middle panel of Fig.~\ref{histogram} represents quality of our data set, i.e. a distribution of mean error ($\rm ERR_{\rm mean}$) of every observed point in the light curve of a star. $\rm ERR_{\rm mean}$ mainly depends on the weather conditions during observations and stellar brightness. Finally, the bottom panel represents a distribution of a standard deviation (STD) of observed light curves. STD is a quality parameter for non-variable stars, but a large STD may also indicate stellar variability if quality of the data is good.

For the initial analysis we used the Muniwin program, which makes a  semi-automatic search for variable stars in a series of source files. An increased value of standard deviation of the observed light curve was used as an indicator of stellar variability. 

To search for detailed amplitude variations we applied other mathematical tools such as the Fourier transform (FT)  (\citealt{Fourier1822}), Lomb-Scargle (LS) periodogram  (\citealt{Lomb76}; \citealt{Scargle82}), or trending parameter (TR) evaluation. 

We assumed that a star may be variable when a power of signal in the Lomb-Scargle periodogram exceeded a detection limit calculated with FAP=0.01. 

If a star showed continuous short periodic variations of brightness, we used the Period04 code  which calculates FT spectra and performs a least squares fit of a sinusoidal function to the observed light curves, which are time dependent functions $f(t)$, 
in order to find frequencies $\Omega$, amplitudes $A$ and initial phases $\Phi$ of the pulsations (\citealt{Lenz05}):
$$f(t)=Z+\sum A_i \sin (2\pi(\Omega_i t+\Phi_i))  $$
Here $Z$ is a zero point of the time dependent function and usually is equal to the mean apparent magnitude of the variable star. As zero point $Z$ does not have influence on variability parameters such as  $\Omega$, $A$, and $\Phi$ we normalized the light curves to $Z=0$, i.e. for the analysis we used the observed relative magnitudes of the variable stars.

In the case of eclipsing binaries, the LS periodogram or FT spectrum does not exhibit a significant peak of variations, however, such variables were easily recognised from typical shapes of their light curve (LC), which allowed us to measure depths and duration of eclipses, and a possible orbital period. 

For long periodic variables which showed just a slow one way change of magnitude during our observations, we calculated a trending parameter (TR$_{\rm O}$) using the following equation: 
$${{{\rm TR_O}=\frac{m_{\rm max}-m_{\rm min}}{\rm STD_{LC}\cdot ERR_{\rm mean}}}}, $$
where $m_{\rm max}$ and ${m_{\rm min}}$ are differential magnitudes determined at phases when a star was faintest and brightest, respectively. In our case, we divided our observed LC into six chunks, calculated values of the mean differential magnitude for every chunk and used only maximal and minimal values for the TR calculations; 
${\rm STD_{\rm LC}}$ is a standard deviation (STD) of LC; 
${\rm ERR_{\rm mean}}$ is a mean value of measurement errors.

 \begin{figure}[t]
   \includegraphics[width=\hsize]{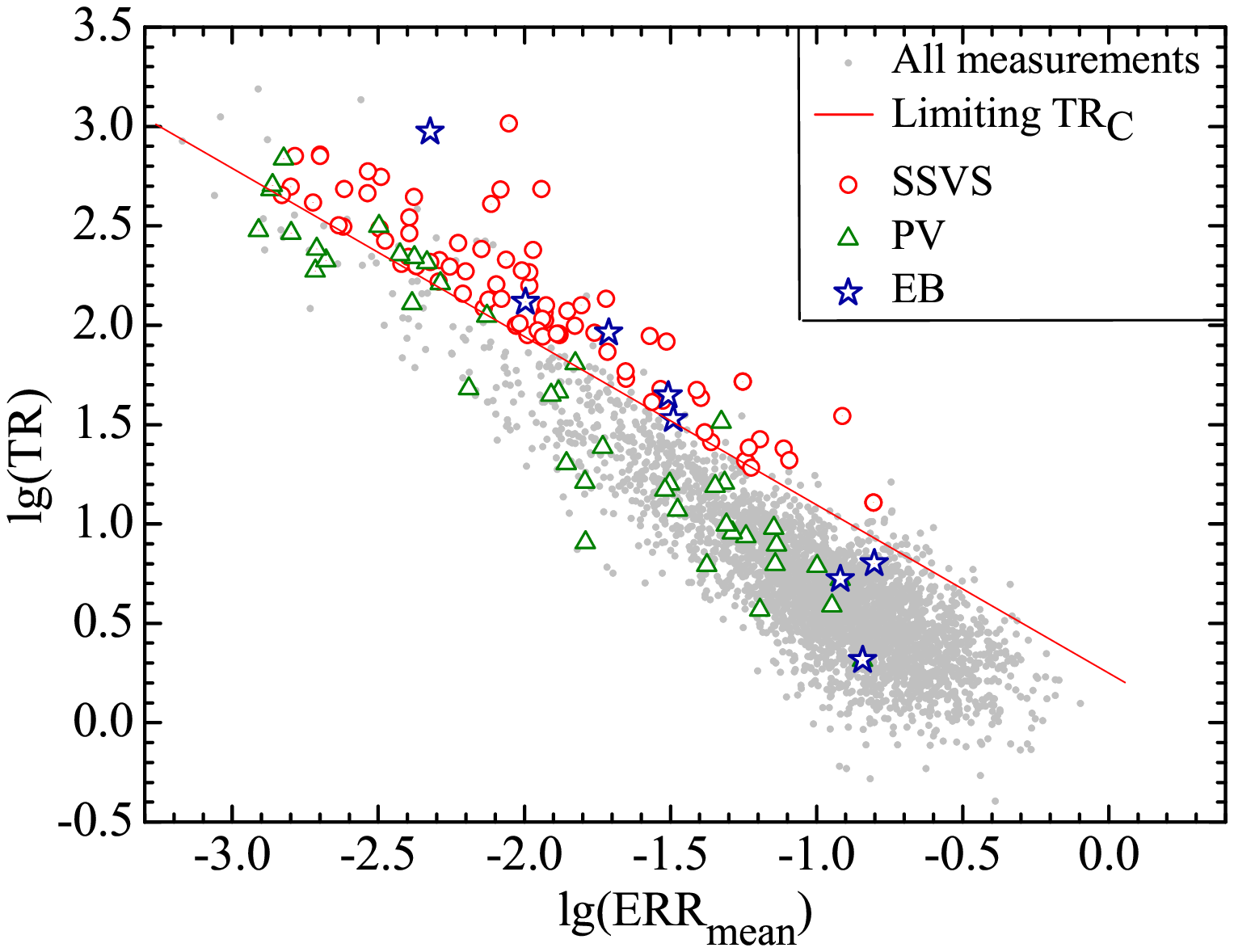}
      \includegraphics[width=\hsize]{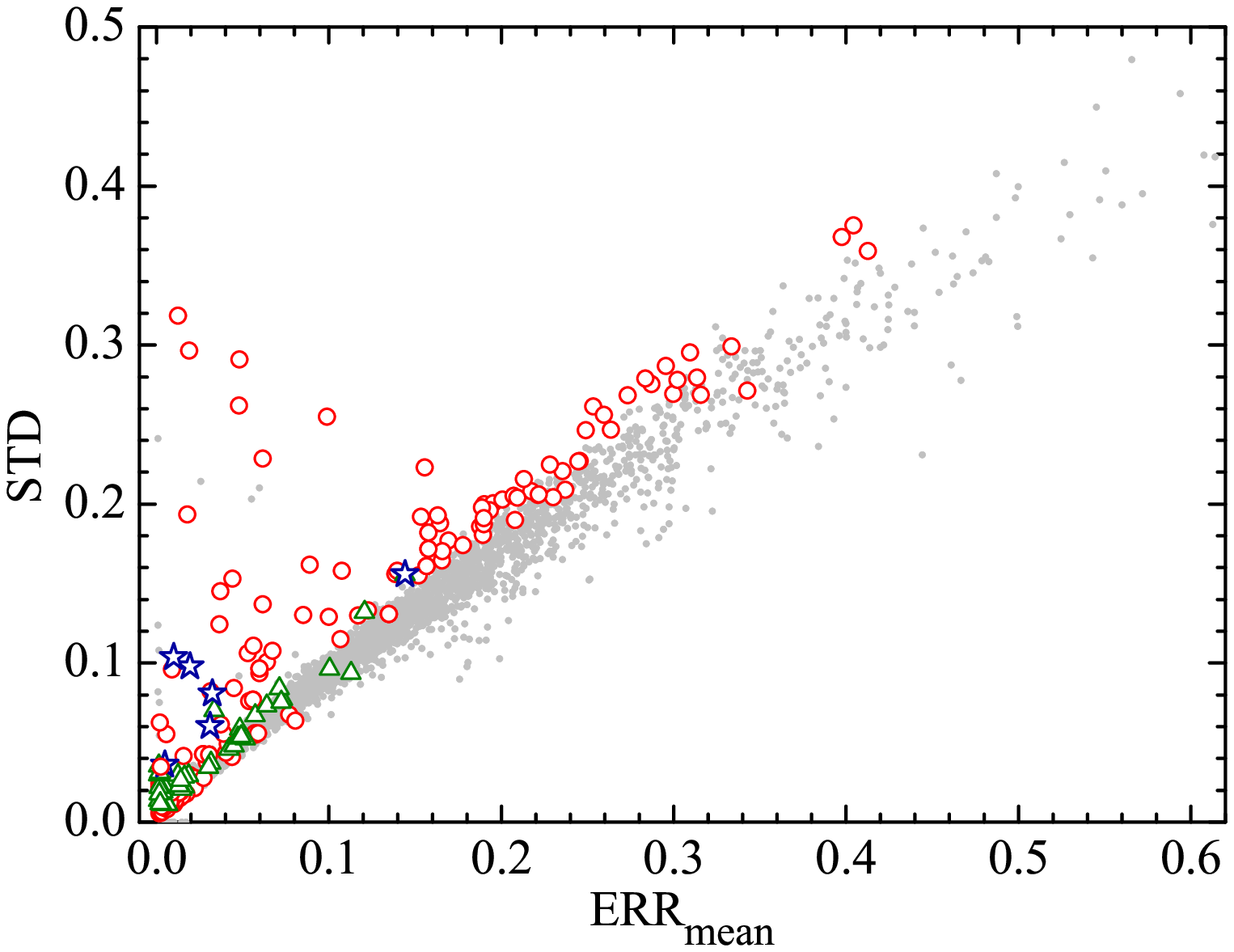}
\caption{Dependencies of TR and STD parameters on the mean error of measurements. See the text for more explanations. 
              }
\label{TR_STD}
   \end{figure}

Fig.~\ref{TR_STD} represents different types of the observed variable stars or candidates for variable stars in lg(TR) vs. lg(ERR$_{\rm mean}$) and in STD vs. ERR$_{\rm mean}$ diagrams.   
The continuous line in the upper panel corresponds to a limit of trending parameter above which slowly varying stars are normally located. The TR limit in logarithmic scale $\lg\rm (TR_C)$ was calculated as a linear equation:
$$\lg(\rm{TR_C})=0.251525-0.84602865\cdot lg(ERR_{\rm mean})$$ 

Majority of non-variable and 
short-periodic variable stars lay below the limit of trending parameter,
however, some stars are above with the same measurement error. The stars above the limit  may have trending or slowly varying LCs. 

We computed the difference between the observed $\rm \lg(TR_O)$ and calculated $\rm \lg(TR_C)$ at a certain $\rm ERR_{mean}$:
$$\rm lg(TR_{O/C})=lg(TR_O)-lg(TR_C)$$
and checked all the stars with positive $\lg\rm (TR_{O/C})$ since they may be potential variables. 

The lower panel in Fig.~\ref{TR_STD} is also commonly used for identification of variable stars. 
Every star which appears above the distribution of non-variable stars may be 
a variable star.  
 We also find other stars there with larger STD than normally, but they do not show any long or short periodic variability. Most of those stars have a close neighbour on the sky, their LC is affected by the light of the neighbouring star and STD of the LC gets increased in comparison to solitary stars. This effect becomes particularly strong when a neighbouring star is much brighter and the sky has thin clouds or haze. We checked all these cases.
 
The calibrated CCD images (24\,470 images in total) and various resulting files were collected into a catalogue of analysed stars and are available online at the Lithuanian National Open Access Research Data Archive (MIDAS)\footnote{https://www.midas.lt  DOI:XXX.XXXX.XXXX}.  

The CNAMEs for the stars in our survey were constructed according to their coordinates. For example, the star with the coordinates  
$\alpha_{2000}$=23:28:23.6374 and $\delta_{2000}$= +19:53:05.969 has the CNAME 23282364+ 19530597.

\section{Results}

We found 81 
new variable candidates in the observed fields and analysed their light curves. We found one eclipsing binary star,  ten variable stars with periods between 35 minutes and 20 days, and 70 slowly varying stars of so far undefined periodicity. Variability analysis was also done for the eclipsing binary candidate TYC\,2764-1997-1 (both from photometry and the observed spectra), and for three known eclipsing binary stars  ASASSN-V J011509.99+652848.0, J213047.63+161528.6, and J215042.53+193829.5. 
In the on-line data catalogue we present light curves of 23 already known variables: 13 $\delta$\,Scuti stars investigated in our previous work (\citealt{Pakstiene2018_deltaScuti}), for the the rapidly varying V*\,RZ\,Psc, the classical Cepheid V*\,BP\,Cas, the high mass X-ray binary V*\,V662\,Cas, the rotationally variable star BD\,+70\,824, for the long-periodic~~ variables~~ IRAS\,21284+1620,~~ 2MASS J21293856 +2440506, and 2MASS J21294742+1646380, for~~ the~~ eclipsing~~ binaries~~ of~~ W\,UMa~~ type~~ CRTS J215111.6+192342 and CRTS\,J022418.8+413147, and for the eclipsing binary of Algol type CRTS\,J231805.7+ 361253).

The results obtained for individual variable stars analysed in our study are described below. 

\subsection{Eclipsing binary candidate TYC\,2764-1997-1}

First of all, we present our results for TYC\,2764-1997-1 (23194862+36034925  in our catalogue) which was marked as an unconfirmed eclipsing binary in the SIMBAD database.  
For the first time it was catalogued as a contact binary candidate in \citet{Gettel2006}. \citet{Hoffman2009} classified TYC\,2764-1997-1 as a double star with a period of 7.8012 hours and an amplitude up to 0.285~mag, as derived using observations in the Robotic Optical Transient Search Experiment (ROTSE-I). 

We had eight photometric runs and some sporadic measurements of TYC\,2764-1997-1 brightness. 
The entire LC and FT spectrum are presented in an on-line catalogue. Analysis of the FT spectrum gave the main period of brightness variability of 3.90~hours with an amplitude of 160.39~mmag, which should be the mean period of the eclipses. 
Analysis of the phased light curve gave the orbital period equal to 7.8028 hours.

  \begin{figure}
   \centering
   \includegraphics[width=\hsize]{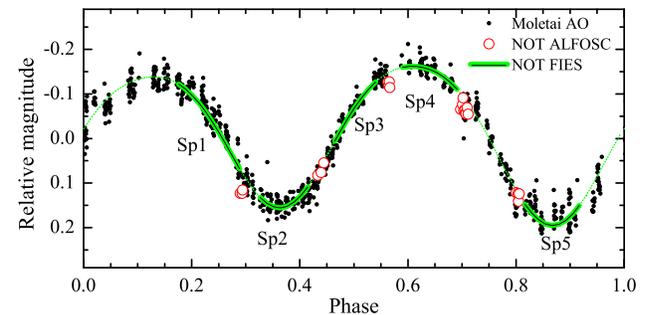}
\caption{Phase diagram of TYC\,2764-1997-1 calculated using the period of 7.8028 hours.  
              }
\label{contact_EB_phase}
   \end{figure}    

On 22 November, 2017, TYC\,2764-1997-1 was also observed at the NOT.  We performed quasi-simultaneous observations with the ALFOSC CCD camera and the FIES spectrograph. Photometric observations with ALFOSC were carried out in order to trace the phase of the brightness variations (Fig.~\ref{contact_EB_phase}) and this confirmed that the orbital period derived using photometric observations at MAO was precise,  whereas the value of orbital period derived by \citet{Hoffman2009} gave a significant shift in phase. 

Fig.~\ref{contact_EB_phase} shows a phase diagram calculated using the orbital period of 7.8028 hours as derived in this work. As we obtained a larger value for the orbital period than \citet{Hoffman2009}, there is a possibility that the orbital period of the system has changed since 2009 either due 
to the redistribution of matter between the stars, or due to the
loss of angular momentum. In case of the conservative
mass exchange, the period increases if mass is transferred
from the less massive to the more massive star (\citealt{Skelton2009}).
A shape of the phase diagram is typical of diagrams of contact binary stars. 
The difference between the primary
and secondary minima is around 0.039~mag (usually this difference is no more than 0.1 or 0.2 magnitudes (\citealt{Skelton2009}).
The difference between maximum and
minimum is 0.3558~mag (typically it is of the order of 0.75 mag (\citealt{Skelton2009}).

The out-of-eclipse brightness maxima in the light curves of TYC\,2764-1997-1 have 
slightly different magnitudes (approximate difference is equal to 0.0239~mag). 
This may be indication of a strong magnetic field at least in one of the components, 
as some W\,UMa-type EB stars may be magnetically 
active and may have spots on their surface. 
This effect is known as the O'Connell effect. As W\,UMa-type stars are circular orbit binaries, 
a spin period of each component  
equals to the orbital period. Such systems  
are very fast rotators and may have strong magnetic fields. The faster the star rotates, the stronger field it can generate.

Fig.~\ref{contact_EB_phase} also shows a larger scatter of points in approximately 
quarter of the orbital period  (from the second minima up to the first maxima). This may indicate instability of one of the components or spots on the surface. 
    
  \begin{figure*}
   \centering
   \includegraphics[width=17cm]{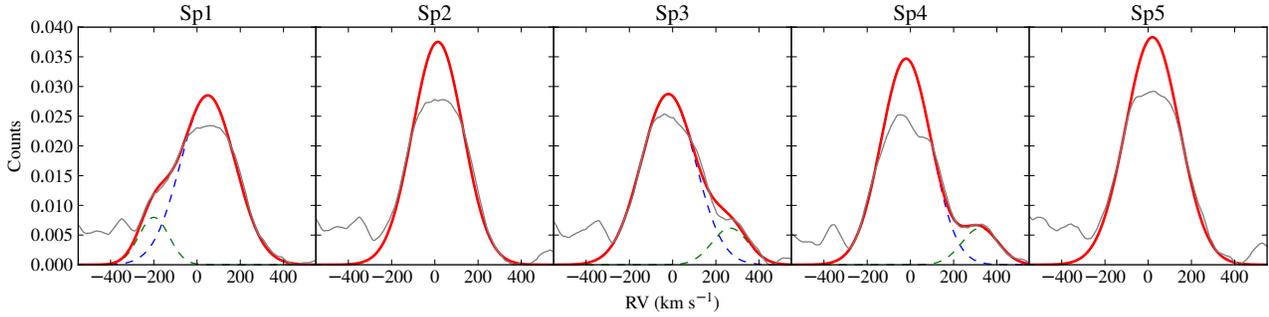}
\caption{CCFs produced for calculating  radial velocities of the 
double-line eclipsing-binary star  TYC2764-1997-1 (grey solid line). Gaussian fits in CCFs of Sp1, Sp3, and Sp4 spectra with a double component detectable are shown as dashed green and dashed blue lines for weaker and stronger components, respectively, whereas the solid red line is a superposition of both fitted Gaussians. For CCFs of Sp2 and Sp5, only one component was fitted and marked as the solid red line.
              }
\label{contact_EB_CCF}
   \end{figure*}     
   
  \begin{figure*}
   \centering
   \includegraphics[width=17cm]{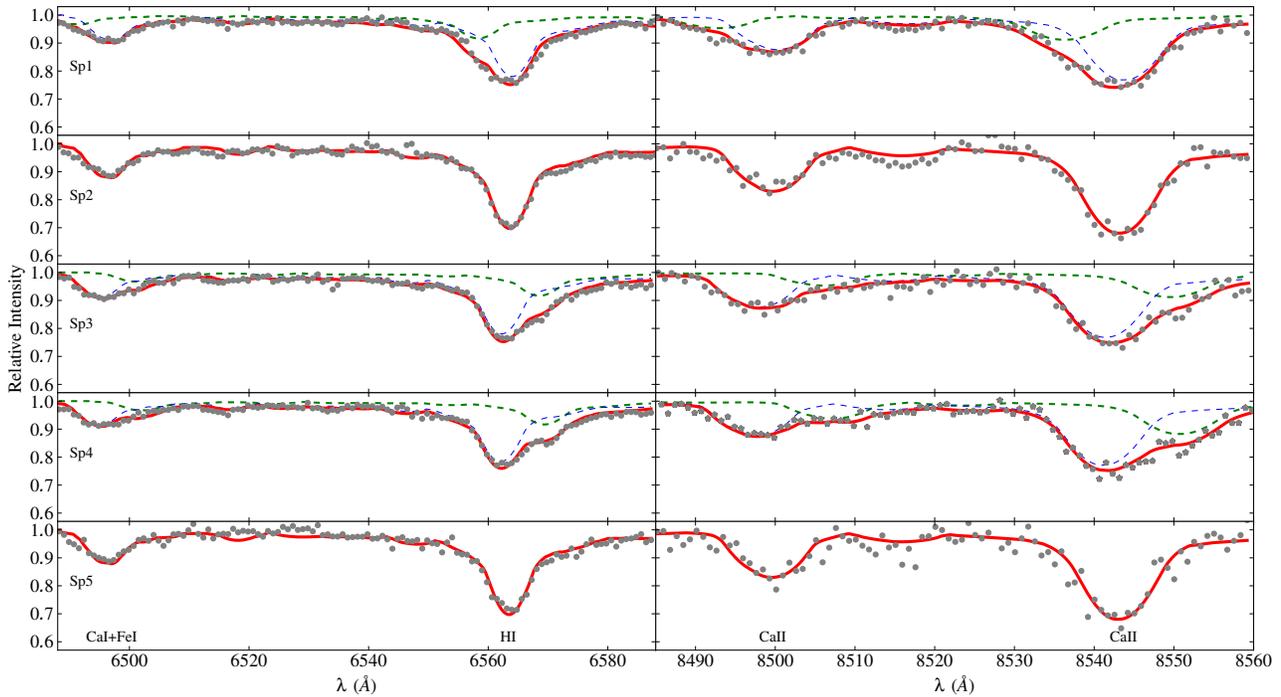}
\caption{Examples of the five TYC2764-1997-1 (grey dots) spectra. Synthetic spectra computed with Solar parameters, $v$sin$i$=180~km~s$^{-1}$  are approximately fitted using radial velocity values from Fig.~\ref{contact_EB_CCF}. The dashed green and dashed blue lines are of  weaker and stronger components, respectively, whereas the solid red line is the sum of both synthetic spectra. For CCFs of Sp2 and Sp5 only one component was fitted.
              }
\label{contact_EB_SP}
   \end{figure*}

On the same night November 22, 2017, we obtained five high resolution spectra with the NOT FIES spectrograph.  The obtained spectra have a high $R=$ 67\,000 resolving power and cover a spectral range from 3700~to~8300~\r{A}. Signal-to-noise ratios (S/N) for an individual spectrum at 6500~\r{A} are 20--25. Thus, in order to increase S/N, the original spectra were rebinned by the factor of 4.

The spectra have shown that the object is a fast rotator with highly broadened lines of absorption. Comparison of the spectra obtained at two consecutive minima of LC showed a difference of the system color, which indicates that stars of different type dominate at every second observed minimum of LC (a usual effect in eclipsing binary systems). Moreover, H$_{\alpha}$ line becomes split into two lines of different depths at brightness maximum. Therefore we can confirm that TYC\,2764-1997-1 is an eclipsing binary system.

Then we performed a cross-correlation of spectra with a synthetic mask based on the atomic line list in order to understand this spectroscopic binary system. A line list for the mask (258 atomic transitions of 25 chemical elements) was composed choosing strongest transitions of the Solar spectrum according to the atlas of the Solar spectrum (\citealt{Wallace2011}). For each spectrum a cross-correlation function (CCF) ranging from $-600$ to +600 km\ s$^{-1}$ was calculated in radial velocity steps of $\Delta$V$_{rad}$=1.2~km\ s$^{-1}$. Figs.~\ref{contact_EB_SP}~and~\ref{contact_EB_CCF} show examples of the spectra and their CCFs. We plotted synthesised spectra on top of the observed spectra (see Fig.~\ref{contact_EB_SP}) in order to test the derived radial velocities. The line-profiles in the spectra and CCFs are widely broadened due to two reasons. Firstly, both stars are fast rotators. Secondly, due to long exposure time (2500~s) that covers a significant part of the phase, the line profiles should shift during the exposure. The line profiles and corresponding CCFs are not perfectly Gaussian due to this reason. Thus, the measured radial velocities should be treated as mean values of the part of the phase covered by exposures of 2500~s.

\addtolength{\tabcolsep}{-3pt}
\begin{table}[t]
\caption{Radial velocity estimates of two TYC\,2764-1997-1 components in km\,${\rm s}^{-1}$.}              
\centering                          
\begin{tabular}{c c c c c c}        
\hline\hline                 
Sp. & JD$-$2457500&$V_{\rm rad}$       &$\sigma_{V_{\rm rad}}$ & $V_{\rm rad}$         & $\sigma_{V_{\rm rad}}$ \\
\hline
   & &\multicolumn{2}{c}{Component~1} &\multicolumn{2}{c}{Component~2}\\
Sp1 & 580.31395 & $48$ & 126  & $-198$ & 70 \\
Sp2 & 580.35954 & $24$ & 117    & & \\
Sp3 & 580.40609 & $-20$ & 125   & 266 & 84 \\
Sp4 & 580.44879 & $-29$ & 115 & 312 & 84  \\
Sp5 & 580.52316 & $21$ & 125  & & \\
\hline                                   
\end{tabular}
\label{Table3}
\end{table}
\addtolength{\tabcolsep}{3pt}

Approximate radial velocities were determined (see Table~\ref{Table3}) by fitting double Gaussian profiles to the three CCFs where double features could be resolved and  single Gaussian where only a single feature was resolved (see~Fig.~\ref{contact_EB_CCF}). 
In Table~\ref{Table3} we provide the radial velocity measurement ($V_{\rm rad}$) and 1$\sigma$ value of the Gaussian distribution ($\sigma_{V_{\rm rad}}$) for every component of the double system. However, the precision of the $V_{\rm rad}$ detection ($\sigma_{V_{\rm RV}}$) is often calculated using the method by \citet{Bouchy2005} which takes into account the width and depth of the CCF and the S/N ratio of the spectrum. According to this method, for the brighter component of the double system the $\sigma_{V_{\rm RV}}$ is up to 0.36~km\,s$^{-1}$ and for the fainter component is up to 2.2~km\,s$^{-1}$.
In order to test the derived radial velocities we synthesised a spectrum overplotted on top of the spectra (see~Fig.~\ref{contact_EB_SP}). We adopted the Solar MARCS (\citealt{Gustafsson2008}) model atmosphere, the version v12.1.1 of the spectrum synthesis code TURBOSPECTRUM  (\citealt{Alvarez1998}) and the rotational profile $v$\,sin\,$i$=180~km~s$^{-1}$ for this test.

\subsection{New eclipsing binary star TYC4038-693-1}

\begin{figure}
\centering
   \includegraphics[width=\hsize]{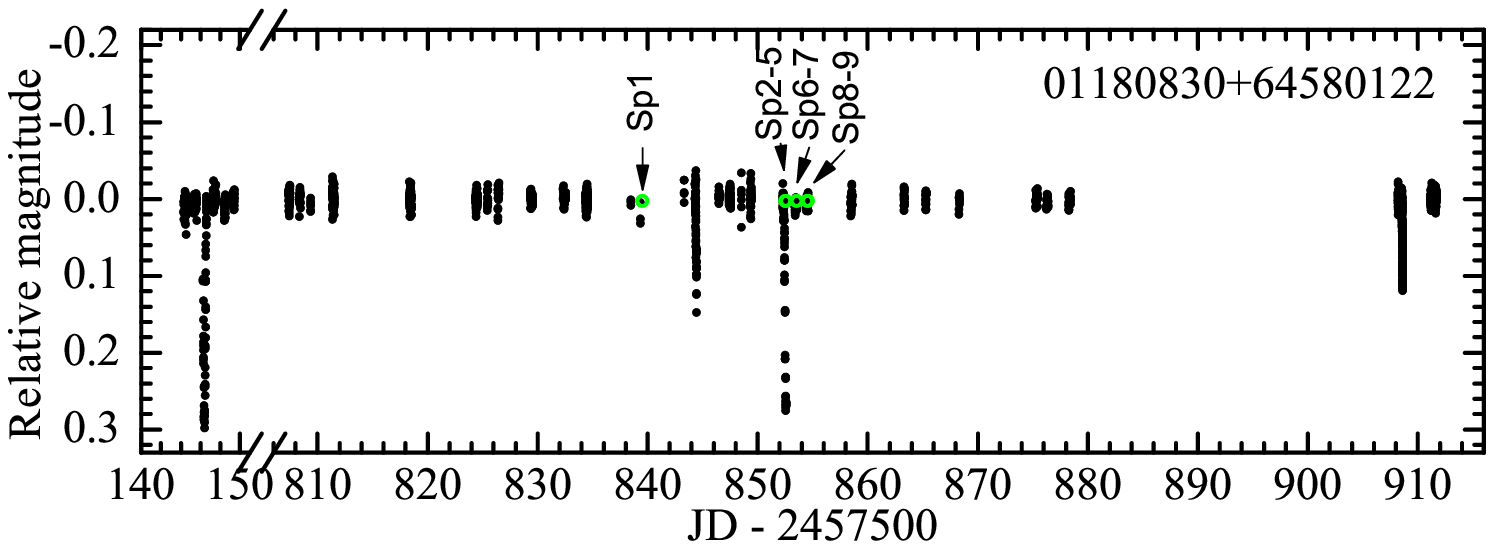}
   \includegraphics[width=\hsize]{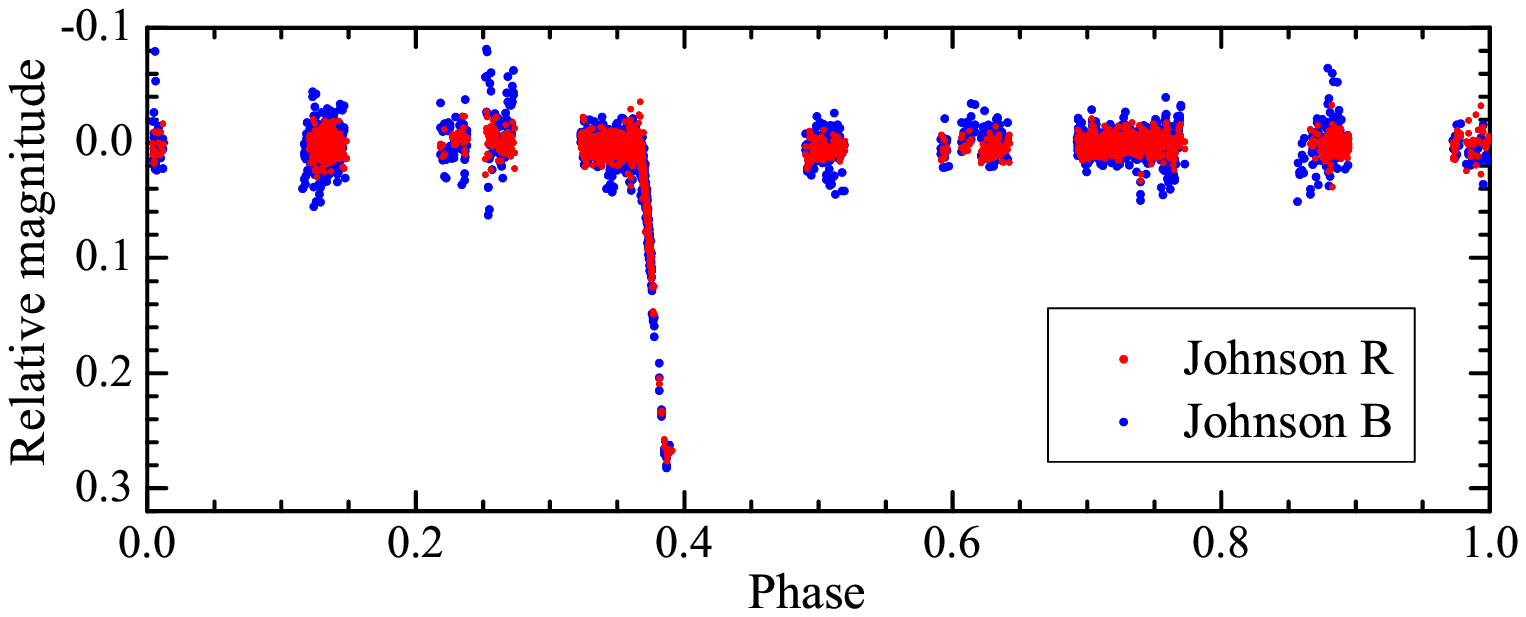}
   \includegraphics[width=\hsize]{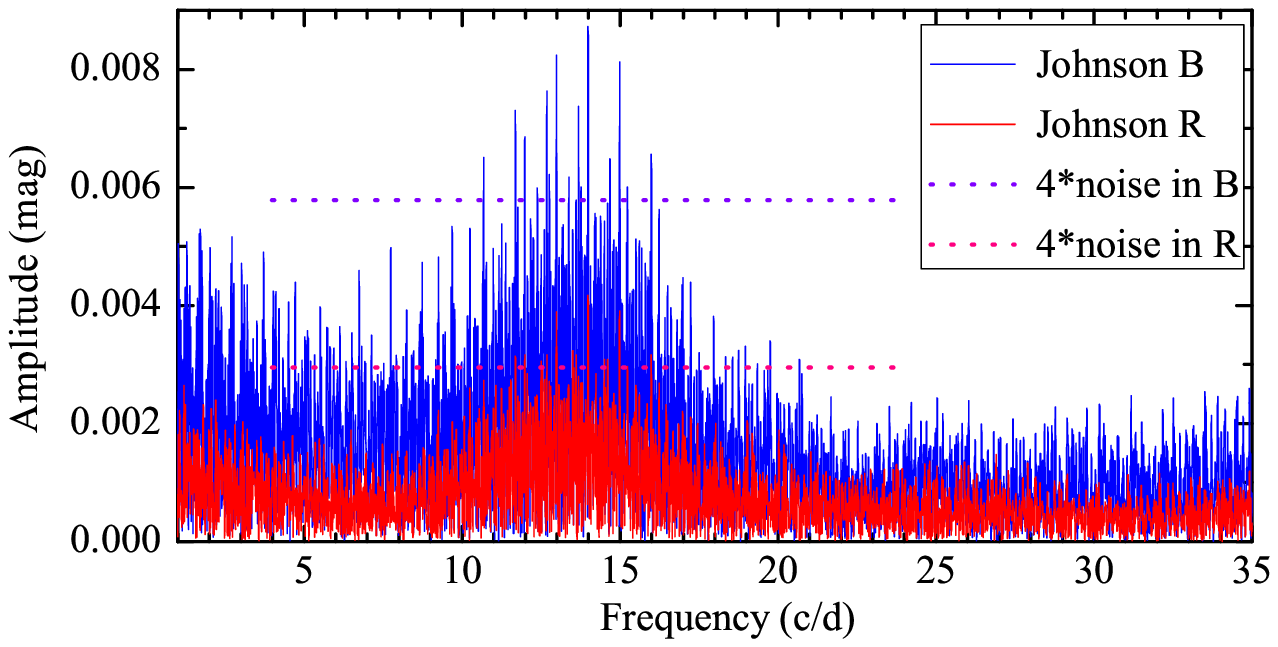}
   \includegraphics[width=\hsize]{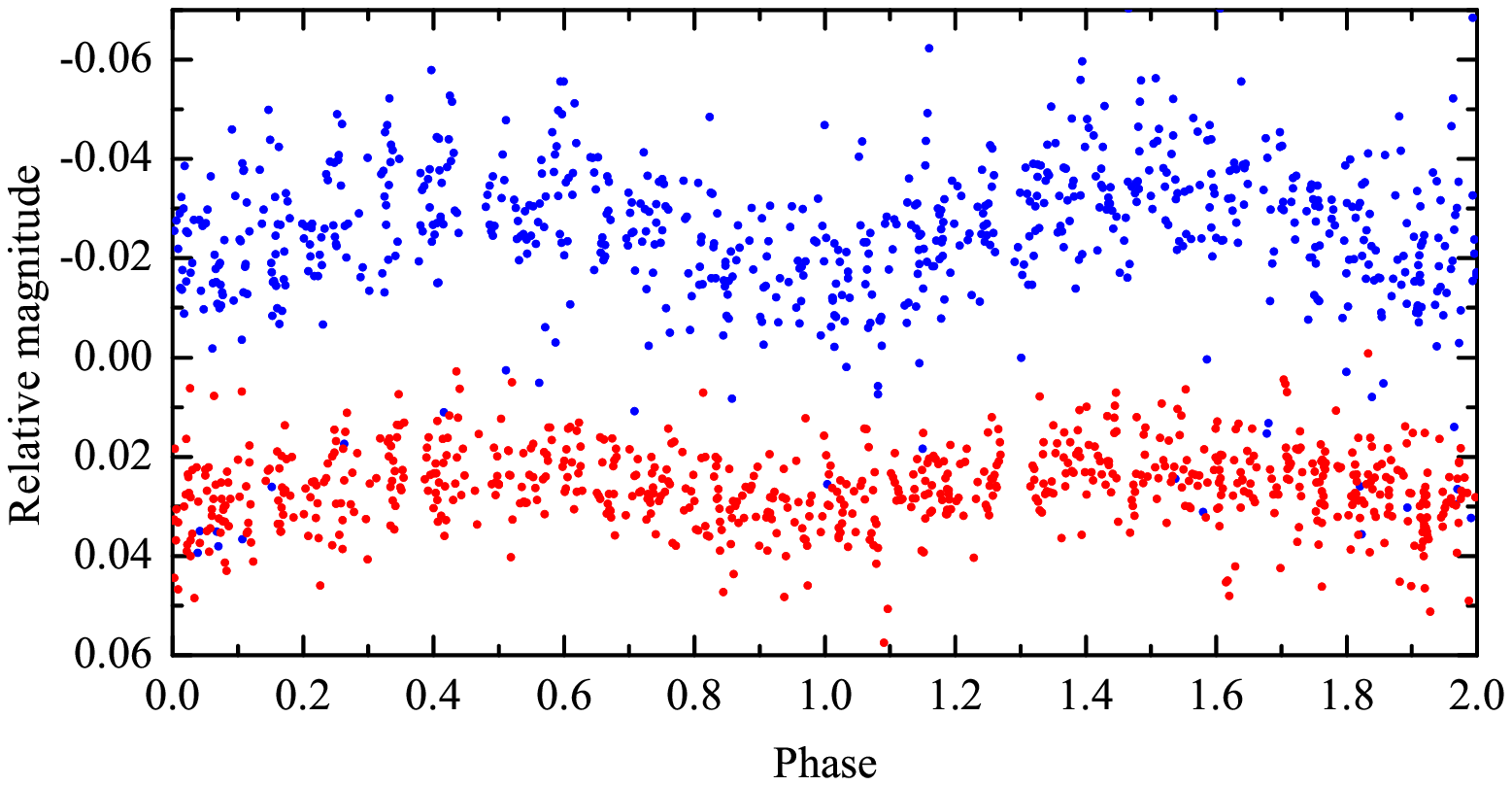}   
\caption{Observed (the upper panel) and phased (the second panel) LCs of TYC4038-693-1 calculated with a period of 8.024~days. The green circles in the first panel show moments when spectra of the system were taken. The third panel shows the amplitude spectra of short periodic variations observed in 2018. The bottom panel shows the phased LCs calculated using the determined period of short periodic variations. }
\label{new_EB_photom}
   \end{figure}

We found TYC4038-693-1 (01180830+64580122 in our catalogue) to be an eclipsing binary star in 2016, when  two partial eclipses were observed (the upper panel of Fig.~\ref{new_EB_photom}). 
As the light curve contained only 7 runs with gaps between them, we 
observed this binary again in 2018 using the Johnson $B$ and $R$ filters in order to characterize the system and determine its orbital period. We observed 31 runs and managed to catch only three partial eclipses with the ingresses (the upper panel of Fig.~\ref{new_EB_photom}). A time gap between the nearest two eclipses was 8.024~days. The second panel of Fig.~\ref{new_EB_photom} shows a phased LC calculated assuming that this time gap might be an orbital period of the system. Depths of the eclipses in $B$ and $R$ filters were almost the same. The depth in the Vilnius $Y$ filter observed in 2016 was deeper approximately by 0.01~mag. 
It seems that the durations of eclipses in 2016 and in 2018 were different. Such differences in eclipse timing variations may be caused by a third body in the system or an apsidal motion.

We found even more changes in the observed light curve of TYC4038-693-1. While the light curve between the eclipses in 2016 was flat, it was not flat in 2018 anymore. In 2018 we found obvious short periodic variations between the eclipses. Analysis of the amplitude spectrum (the third panel of Fig.~\ref{new_EB_photom}) shows significant signals (S/N $>$ 4) at the 1.7159~h period with amplitudes of 8.61~mmag in the $B$ band and 4.16~mmag in the $R$ band. Phased light curves in $B$ and $R$ bands of these short periodic variations are shown in the bottom panel of Fig.~\ref{new_EB_photom}.

In order to determine an orbital period of the system, we decided to obtain radial velocities of the components with the VUES spectrograph at MAO.    
We observed spectra of this binary nine times during four nights (see the green circles in the first panel of Fig.~\ref{new_EB_photom} and Table~\ref{tab:vueslog}). The obtained spectra have a resolving power of around $R = 36\,000$ and cover a spectral range from 4000 to 8800~\AA. S/N of individual spectra were $20-25$. 
In order to increase the S/N, the original spectra were rebinned by the factor of 3. 

    \begin{figure*}
 \begin{center}
   \includegraphics[width=17cm]{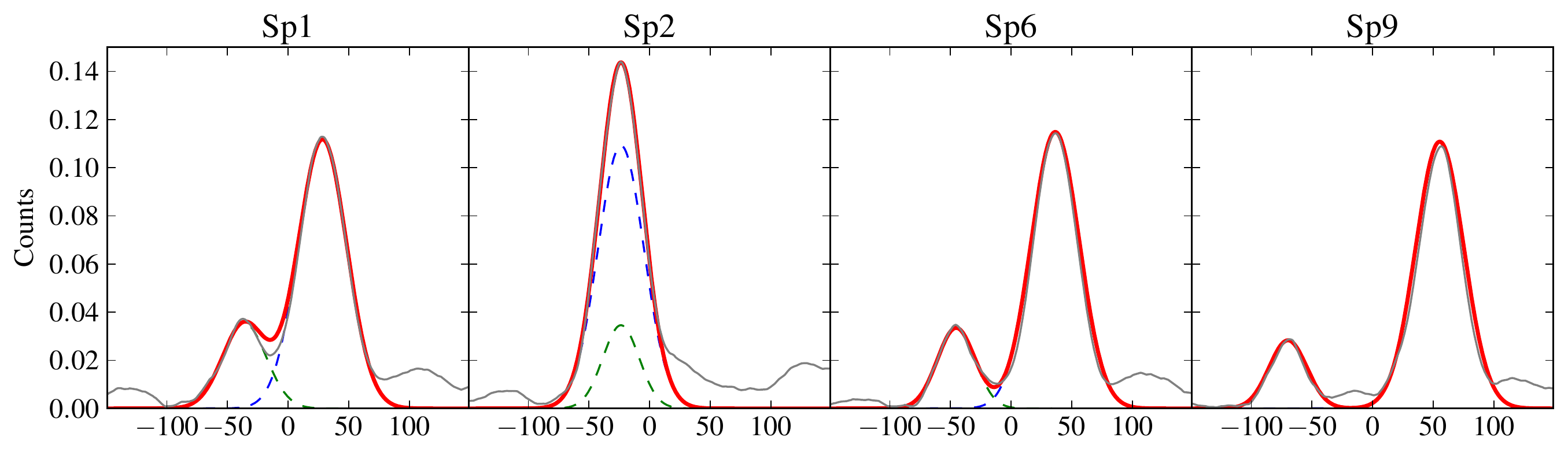}
  \end{center}
\caption{CCFs produced for calculating radial velocities of the double-line eclipsing-binary star TYC4038-693-1 (grey solid line). Gaussian fits in CCFs which had a double component detectable are shown as dashed green and dashed blue lines for weaker and stronger components, respectively, whereas the solid red line is a superposition of both fitted Gaussians. For CCFs where only one component was fitted, the fits are marked by the solid red line.  }
\label{EB1_CCF}
   \end{figure*} 

    \begin{figure*}
 \begin{center}
   \includegraphics[width=17cm]{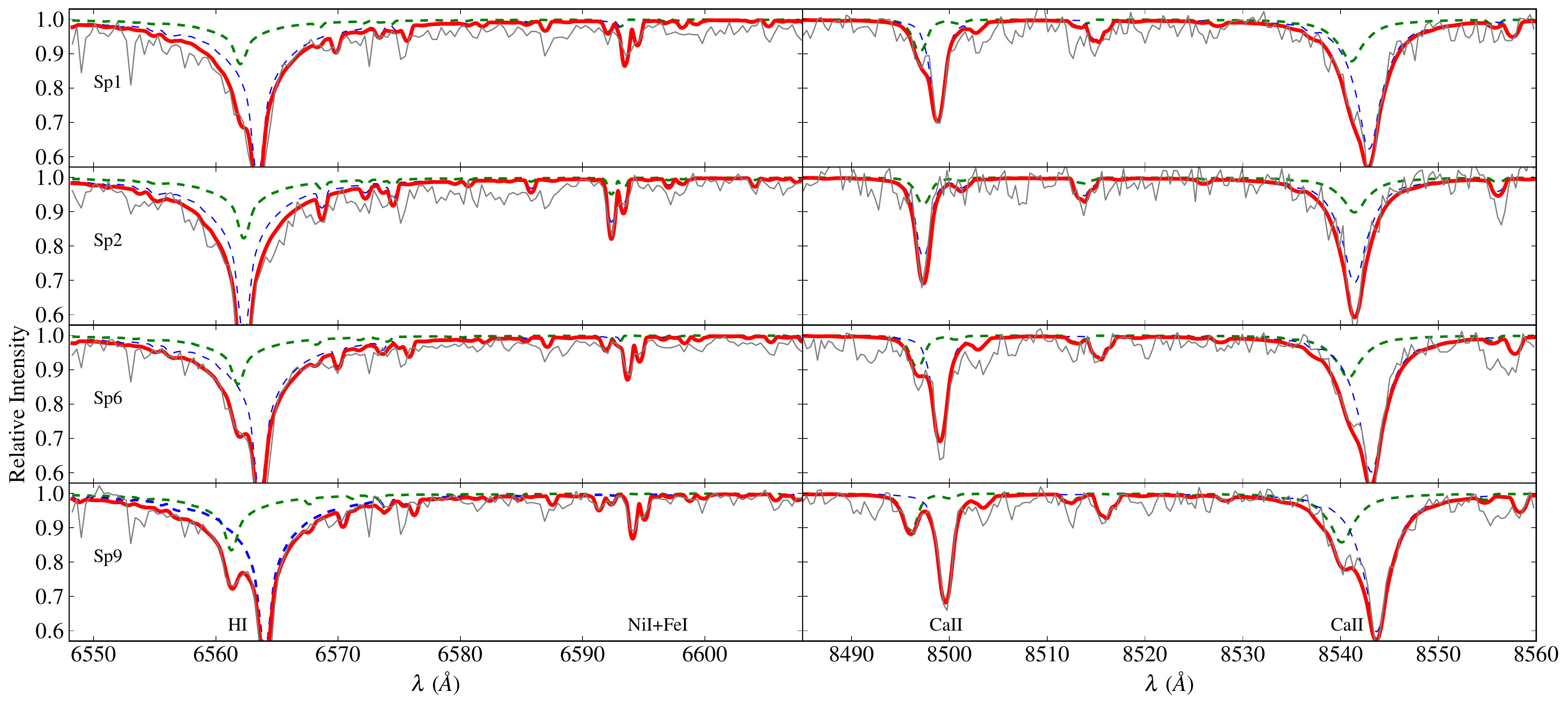}
  \end{center}
  \label{bandymas}
\caption{Examples of spectra of TYC4038-693-1 (grey lines). Synthetic spectra computed with $T_{\rm eff}$=5428.75~K, ${\rm log}~g=4.5$, ${\rm [Fe/H]}=-0.8$, and $v{\rm sin}i$=30${\rm km~s}^{-1}$ are approximately fitted using radial velocity values from Table~\ref{tab:vueslog}. The dashed green and dashed blue lines are for the weaker and stronger components, respectively, whereas the solid red line is the sum of both synthetic spectra. }
\label{fig:MAOSP}
   \end{figure*}

\begin{table*}
\caption{Observing log and radial velocities of two components for the eclipsing binary star TYC4038-693-1}
\begin{tabular}{ccccccc}
\hline\hline
Spectrum & JD$-$2457500 & Exposure time & $V_{\rm{rad}}$ & $\sigma_{V_{\rm{rad}}}$ &  $V_{\rm{rad}}$ & $\sigma_{V_{\rm{rad}}}$\\
   &   &  s  & km\,s$^{-1}$ & km\,s$^{-1}$ & km\,s$^{-1}$ & km\,s$^{-1}$ \\
\hline
&&&\multicolumn{2}{c}{Component 1}&\multicolumn{2}{c}{Component 2}\\
Sp1  &  839.50252& 2400 &  28.40 & 19.33 & $-36.08$ & 17.41  \\
Sp2  &  852.50758& 1200 & $-23.80$ & 17.93 &  &    \\
Sp3  &  852.52188& 1200 & $-23.40$ & 17.94 &  &    \\
Sp4  &  852.53619& 1200 & $-21.43$ & 18.23 &  &    \\
Sp5  &  852.55049& 1200 & $-19.93$ & 18.71 &  &    \\
Sp6  &  853.51569& 2400 &  36.35 & 18.45 & $-45.83$ & 15.14 \\
Sp7  &  853.54388& 2400 &  38.18 & 19.18 & $-47.21$ & 15.02 \\
Sp8  &  854.49791& 2400 &  56.19 & 18.04 & $-69.75$ & 15.19 \\
Sp9  &  854.52609& 2400 &  55.59 & 18.30 & $-69.87$ & 15.63 \\
\hline
\end{tabular}
 \label{tab:vueslog}
\end{table*}

We obtained CCFs in the same way as for TYC~2764-1997-1. Examples of CCFs are shown in Fig.~\ref{EB1_CCF}. The double line features are clearly visible in the spectra of 2018-08-09, 2018-08-23, and 2018-08-24 nights and they are not visible in the spectra of the 2018-08-22 night. Thus we were able to measure the radial velocities for the components by fitting Gaussian profiles.
For this system we found that $\sigma_{\rm RV}$ is up to 0.04 km\,s$^{-1}$ and 0.14 km\,s$^{-1}$ for the brighter and fainter component, respectively.
We provide a summary of observations and radial velocity estimates in 
Table~\ref{tab:vueslog}.     

Similarly as for the TYC~2764-1997-1 star, we plotted synthesised spectra on top of the observed spectra (see Fig.~\ref{fig:MAOSP}) in order to test the derived radial velocities. We adopted  $T_{\rm eff}$=5430~K that was estimated by {\it Gaia}\,DR2 (\citealt{GAIA_DR2_2018}) and chose ${\rm log}~g=4.5$ that is typical for dwarfs. The combination of ${\rm [Fe/H]}=-0.8$ and $v{\rm sin}i$=30\,${\rm km~s}^{-1}$ was the best to fit single neutral isolated iron lines of the 2018-08-24 00:37:34.523 spectrum where the radial velocity spread between the two components is the largest. We were not able to perform a full spectroscopic analysis of the binary system because of lack of isolated iron lines and insufficient accuracy of the effective temperature of this binary in $Gaia$\,DR2.

The spectrum Sp1 observed on the 9 Aug 2018 was taken 4.9~days before the eclipse observed on the 14th of August, and its CCF shape appeared intrinsic to the situation when companions of the binary system are 0.5--1.0~days before or after the eclipse. As periodicity of the eclipses may be 4.012~days, that means the spectrum Sp1 was taken 0.5--1.0~days before the eclipse. Therefore, the orbital period of the system is 8.024~days. This assumption fits well the spectra taken on 23 and 24 Aug (Sp6 and Sp9) after the eclipse observed on 22 Aug. CCFs of Sp1 and Sp6--9 spectra show that the components of the binary system were moving in the same way at those moments: the component with weaker lines was moving toward the observer, and the component with stronger lines was moving away from the observer. The only orbital period which is equal to 8.024~days can explain such profiles of CCFs. 

However, so far it is not known which of eclipses, primary or secondary, were observed.  
We used the derived orbital period of 8.024~days to calculate a phased light curve of the system (second panel of Fig.~\ref{new_EB_photom}).  The phased LC did not show any sign of eclipse at the distance of 0.5~phase off the observed eclipses, as it is usual in the cases when components of an eclipsing binary star have circular orbits. That means that the components of the binary system TYC4038-693-1 may have elliptical orbits with high eccentricity. 
This system requires longer follow-up observations, including spectroscopy, in order to be fully characterized. The results of the TYC4038-693-1 analysis are presented in 
Table~\ref{results-TYC4038-693-1}.

\addtolength{\tabcolsep}{-3.0pt}
\begin{table}[t]
\centering
\caption{Results for the identified eclipsing binary star TYC4038-693-1}              
\begin{tabular}{l l}        
\hline\hline                 
CNAME & 01180830+64580122  \\
$Gaia$ ID&{524895890346462848}\\
$\alpha$(2000) [h m s] &01:18:08.30 \\
$\delta$(2000) [$^\circ$  $^\prime$  $^{\prime\prime}$]  &+64:58:01.22 \\
$Gaia$ $G$ [mag] & 10.46 \\
Orbital periods [days] &8.024\\
Duration [h] &6.5  \\
Depth [mag]&0.28  \\
Frequency of short-period  & \\
variations  [c/d] & 13.986\\
variations [h]      & 1.7159\\
Amplitude in $B$  [mmag]          & 8.6 (S/N$=5.9$)\\
Amplitude in $R$  [mmag]                       & 4.2 (S/N$=5.6$)\\
\hline                                   
\end{tabular}
\label{results-TYC4038-693-1}
\end{table}
\addtolength{\tabcolsep}{3.5pt}

\subsection{Eclipsing binary ASASSN-V J011509.99+652848.0}

The eclipsing binary ASASSN-V J0115 09.99+652848.0 (01151017+65284617 in our catalogue, also NSVS 1689498) was discovered and classified as an EA eclipsing binary by the project ``All-Sky Automated Survey for Supernovae" (ASAS-SN) \citep{Shappee2014} from the LC observed between JD 2457000$-$2458100). Analysis of the LC gave the orbital period of 1.3135561~days with the amplitude of 0.42~mag  \citep{Jayasinghe2018_II}. 
 
The first session of our photometric observations of the eclipsing binary ASASSN-V J0115 09.99+652848.0 was performed in 2016
and repeated in 2018. We used a shorter sampling time for observations than it was used by the ASAS-SN. That allowed us to get more detailed light curves of eclipses. 

\begin{figure}
 \begin{center}
   \includegraphics[width=\hsize]{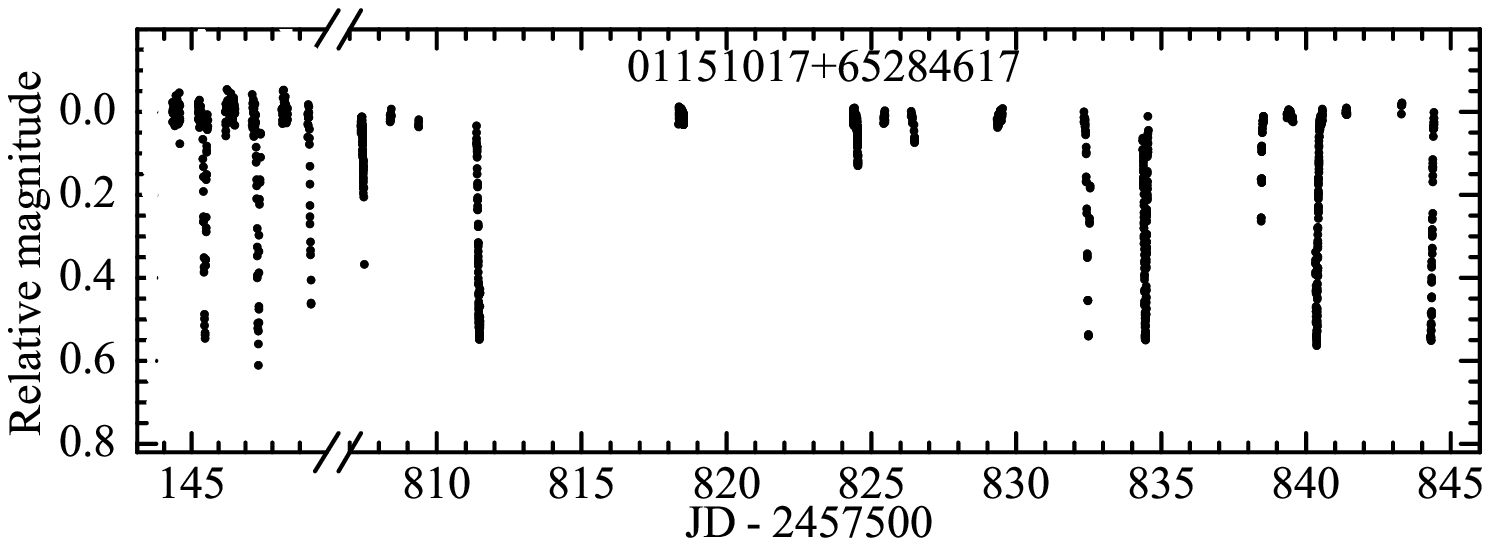}
   \includegraphics[width=\hsize]{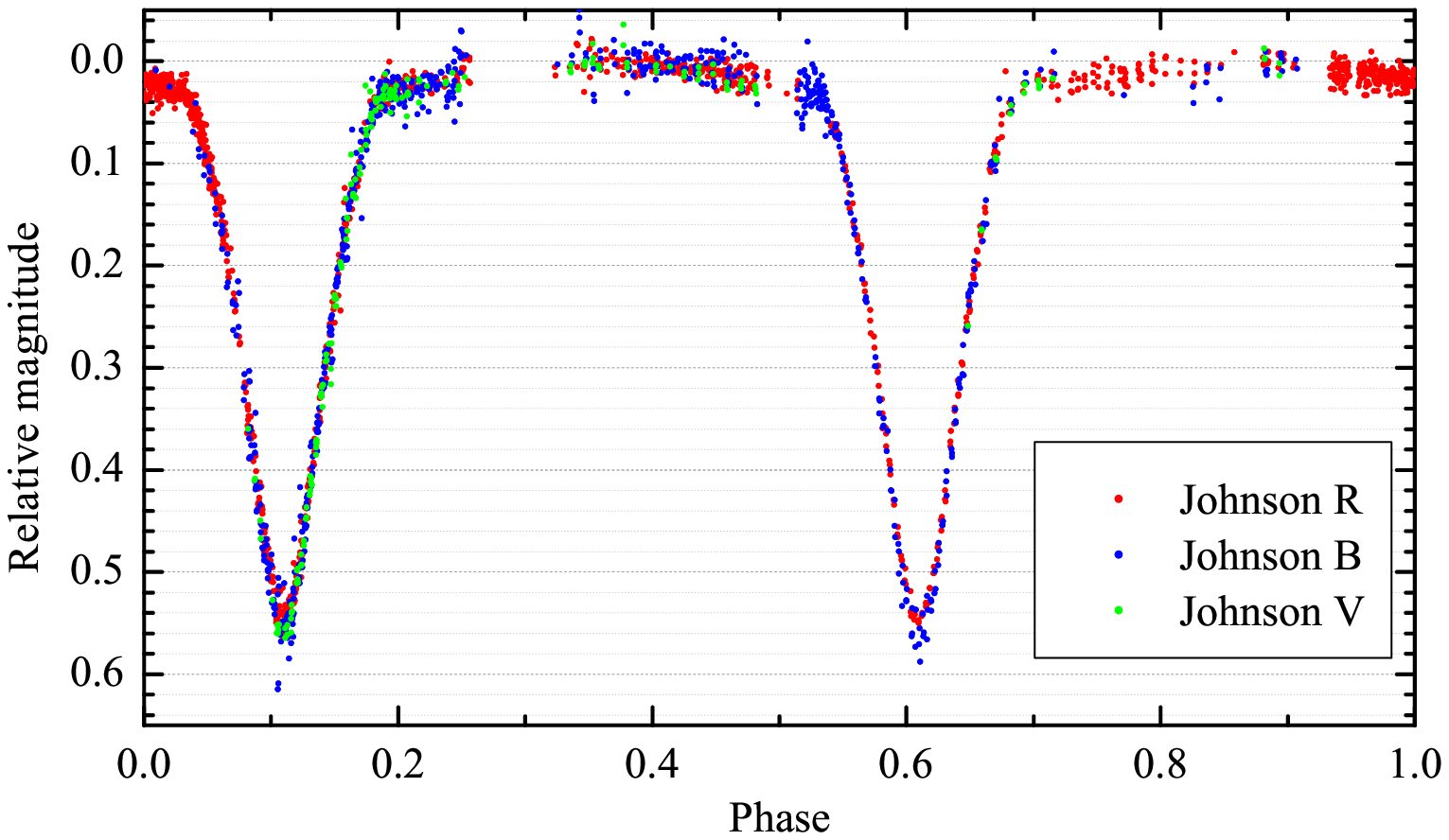}  
  \end{center}
\caption{Observed and phased LCs of ASASSN-V J0115 09.99+652848.0 calculated with the period of 1.31356~days. 
              }
\label{EB2}
   \end{figure}

We observed four eclipses in 2016 with durations of 0.17~days, with sharp bottoms, and the interval of 1.971~days between them (the upper panel of Fig.~\ref{EB2}), but the analysis of the phased light curve suggested four different possible orbital periods. Therefore, we repeated observations of this system in 2018 with the Johnson $B$, $V$, and $R$ filters in order to determine a more accurate orbital period and possible spectral types of the companions (the lower panel of Fig.~\ref{EB2}).  

Analysis of the collected data showed that the system includes two stars with very similar effective temperatures. According to the Gaia DR2 data $T_{\rm eff}$ of the system is 4955~K, which corresponds to the K2 spectral type. The depths of primary and secondary eclipses are almost the same (0.53~mag in $B$ band and 0.51~mag in $R$ band, the bottom panel of Fig.~\ref{EB2}). The light curve between the eclipses is slightly rounded up, which may be indication of tidal effects in the close binary system (prolated shape or hot spots on the surfaces facing one another). Calculation of the phased light curve using observations from 2016 and 2018 allowed us to determine a more precise orbital period of the system, which is equal to 1.31356~days (Table~\ref{threeEB}) and fits well with parameters published by \citet{Jayasinghe2018_II} (orbital period	1.3135561~days with amplitude 0.42~mag).

\subsection{Eclipsing binary ASASSN-V J213047.63+161528.6}
  
The eclipsing binary ASASSN-V J213047.63+161528.6 (21304769+16152983 in our catalogue) was discovered and classified as an EA eclipsing binary by the ASAS-SN project \citep{Shappee2014} from the LC observed between JD 2456450$-$2458100. Analysis of the LC gave the orbital period of 2.20901~days 
 \citep{Jayasinghe2018}.

\begin{figure}
\begin{center}
   \includegraphics[width=\hsize]{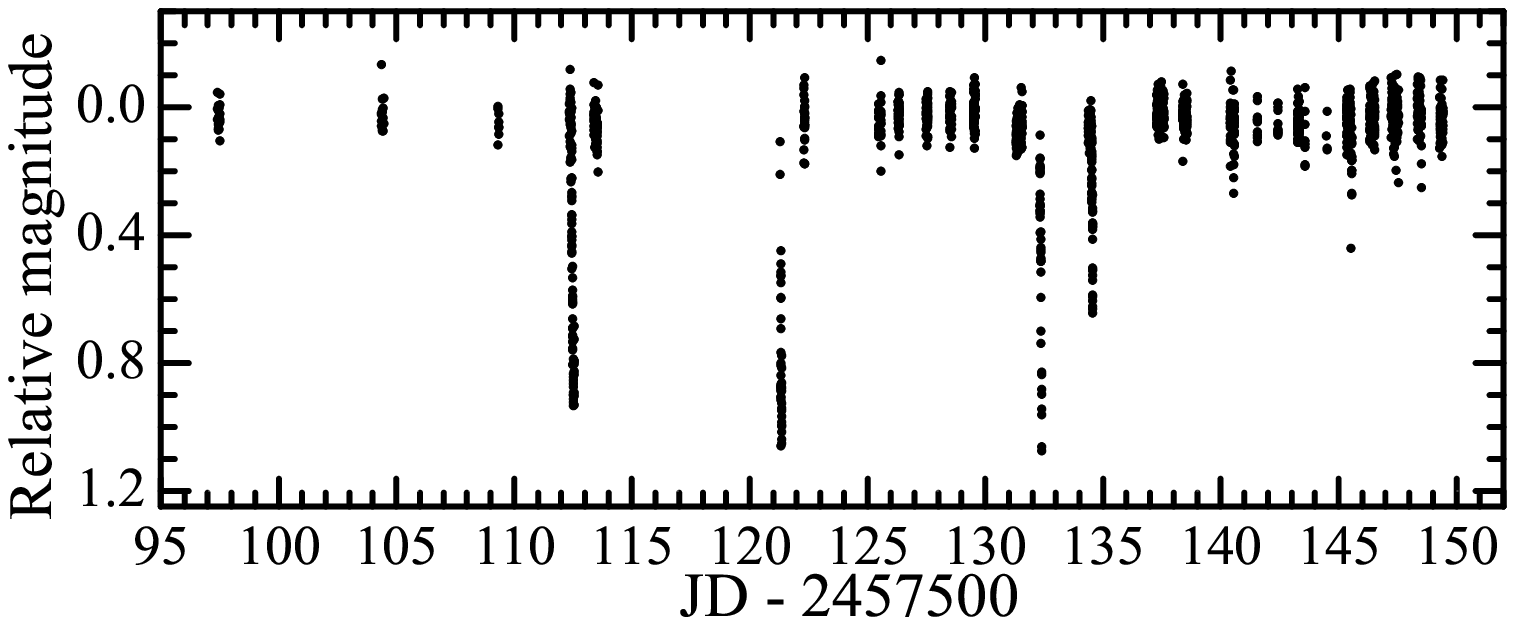}
   \includegraphics[width=\hsize]{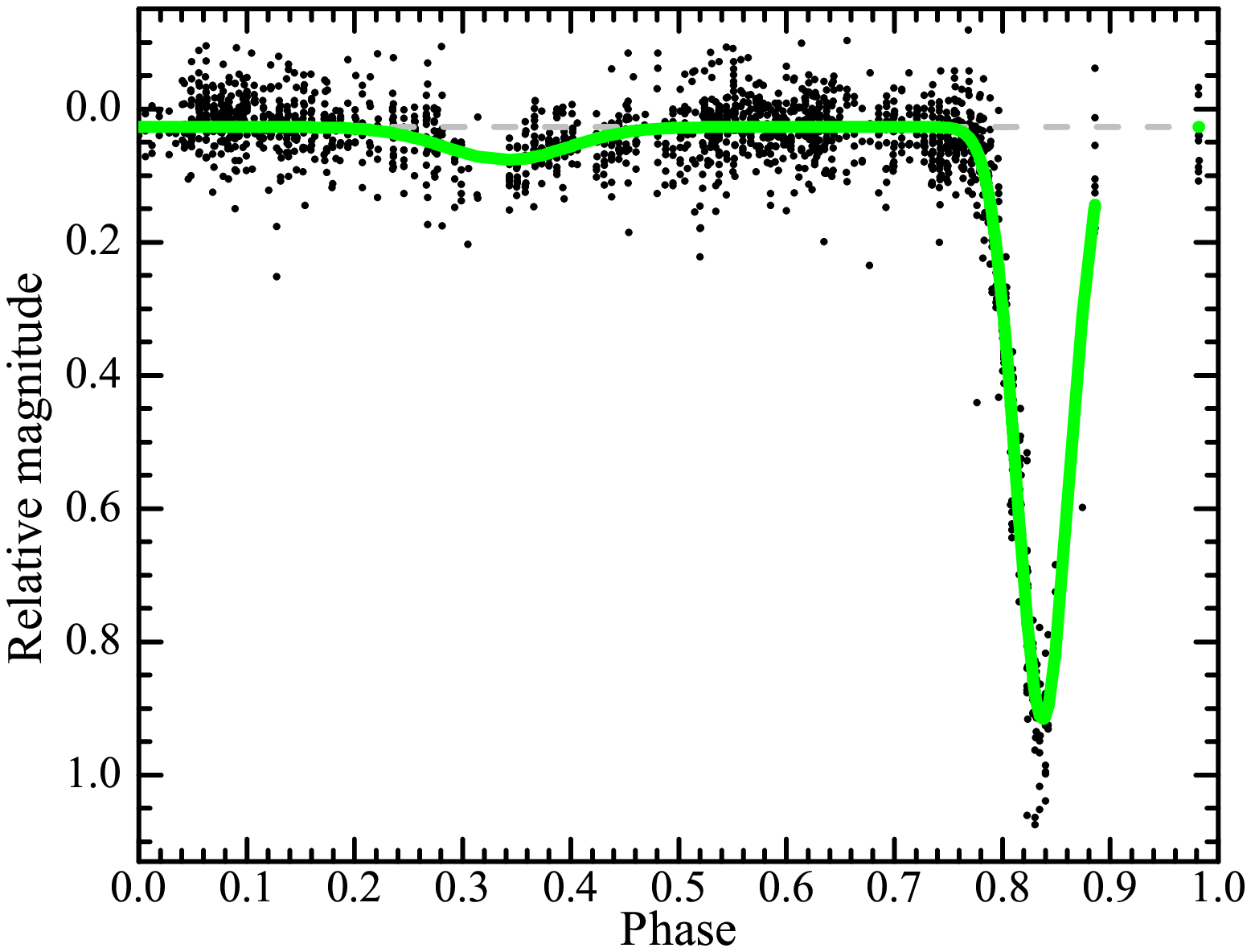}
 \end{center}
\caption{Observed and phased light curves of ASASSN-V J213047.63+161528.6 calculated with the period of 2.21~days. The thick curve corresponds to a Gaussian fit of eclipses in the phased LC, while the dashed horizontal grey line corresponds to a LC without eclipses. }
\label{EB3}
   \end{figure}

We~~ observed ~~the~~ eclipsing~~ binary~~~ ASASSN-V J213047.63+161528.6 in 2016.
The observed LC was quite long, nevertheless, all four eclipses observed were incomplete.
The time between two closely neighbouring eclipses was equal to 2.21~days. 
We used this value to calculate a phased LC, which proved the orbital period of 2.21~days and revealed a secondary eclipse (see Fig.~\ref{EB3}). The orbital period published by \citet{Jayasinghe2018} is also equal to 2.20901~days.

\begin{figure}[t]
 \begin{center}
   \includegraphics[width=\hsize]{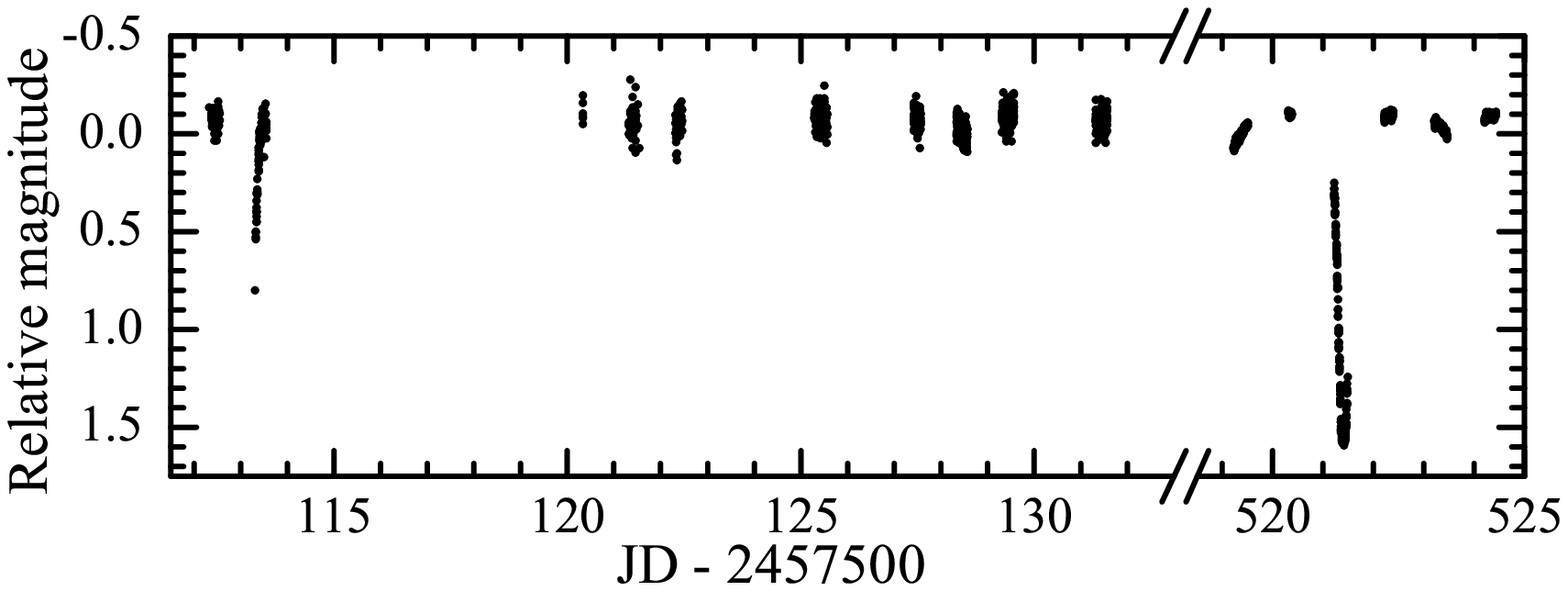}
      \includegraphics[width=\hsize]{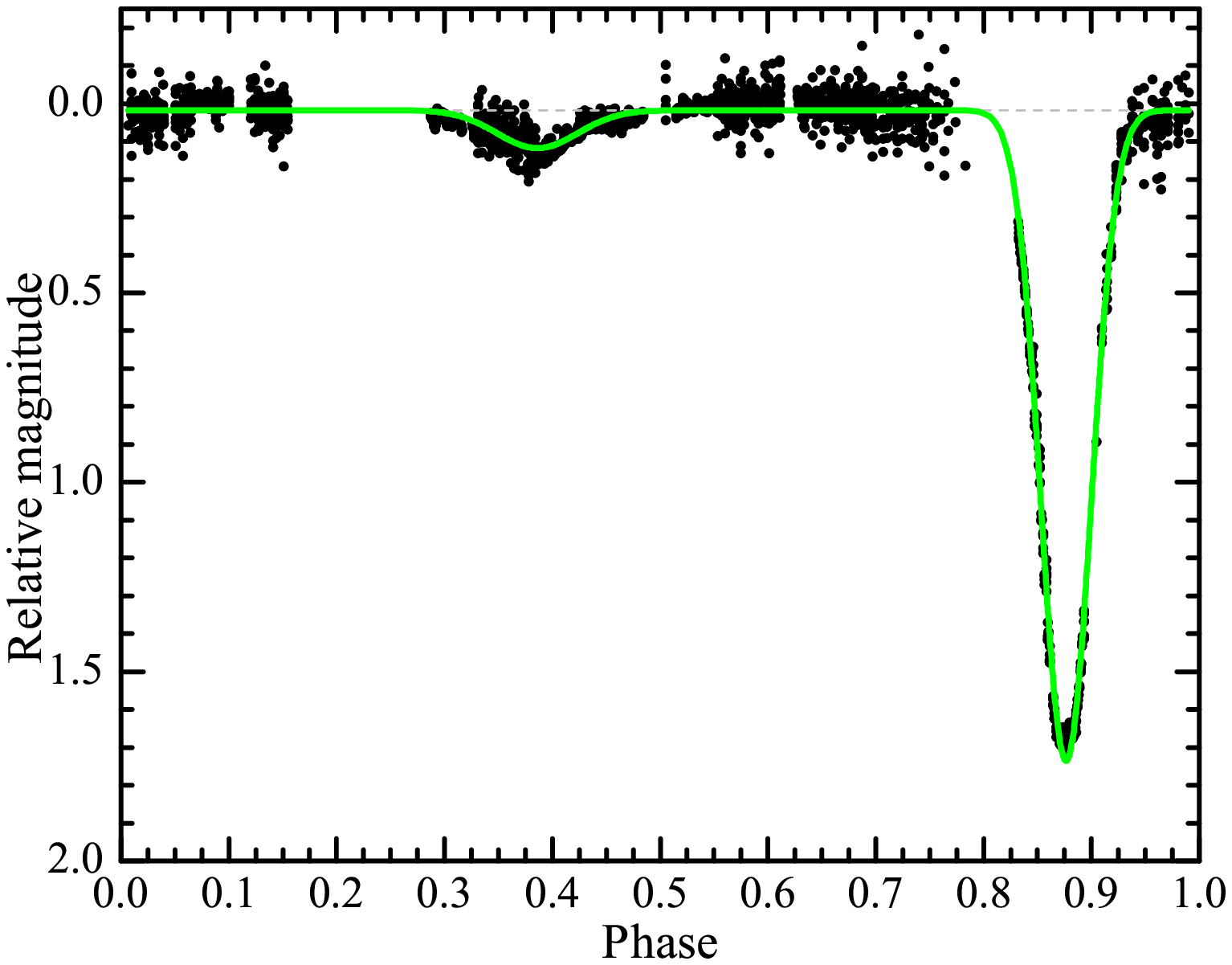}
 \end{center}
\caption{Observed and phased LCs of ASASSN-V J21504 2.53+193829.5 calculated with the period of 4.3896~days. The thick curve corresponds to a Gaussian fit of eclipses in the phased LC, while the dashed horizontal grey line corresponds to a LC without eclipses.
              }
 \label{EB4}
   \end{figure}

 According to the Gaussian fit of eclipses we derived the depths of primary and secondary eclipses, which are equal to 0.882~mag and 0.048~mag, respectively. 
 Duration of the primary eclipse was evaluated according to the moment of the first contact and the middle of the eclipse. As the LC at the moment of the first contact was slightly noisy, we could not determine that moment precisely, which increased the error of primary eclipse duration value, approximately equal to $0.3\pm0.05$~days (Table~\ref{threeEB}). The secondary eclipse was too shallow in comparison to the noise of the LC and we cannot say anything about its duration. 

\subsection{Eclipsing binary ASASSN-V J215042.53+193829.5}

The eclipsing binary ASASSN-V J21504 2.53+193829.5 (21504261+19382841 in our catalogue, also ASAS J215042+1938.5) was discovered and classified as an EA eclipsing binary by the ASAS-SN project \citep{Shappee2014} ~~from~ the~ LC~ observed~ between ~~ JD\,2456250$-$2458100. Analysis of the LC gave an orbital period of 4.389628~days \citep{Jayasinghe2018_II}. 

The first session of our photometric observations of eclipsing binary ASASSN-V J21504 2.53+193829.5 was performed in 2016.
When we observed this star during 12 runs in 2016, only one run showed a typical LC shape of egress with a clear turn point. 
Thus we repeated the observations in 2017 and managed to observe more than a half of the primary eclipse with a clearly detectable flat bottom, but without its starting and ending moments, and two signs of the secondary eclipses on both sides of the primary eclipse (see the upper panel of Fig.~\ref{EB4}).

We used several different values of periods for the phased LC calculation using all our observed data from 2016 and 2017 and found that the best fitting period value is 4.3896~days. It fits well with the parameters published by \citet{Jayasinghe2018_II} (orbital period 4.389628~days). This period gave the phased LC shown in Fig.~\ref{EB4}. A closer look at the phased LC has shown that LCs of the secondary eclipses observed in 2016 and 2017 do not fit to each other exactly, while the primary eclipses form an almost continuous depression/pit. 

 Using the Gaussian fit of eclipses we derived depths of the primary and secondary eclipses, which are equal to 1.665~mag and 0.160~mag, respectively (\citet{Jayasinghe2018_II}, gives the amplitude equal to 1.52~mag).  The duration of the flat part between the second and third contacts of primary eclipse is 0.4136~hours.  
As we did not observe the moment of the first contact, we assumed that the duration between the first and the second contacts is equal to the duration between the third and fourth contacts. Thus, the total duration of the primary eclipse was calculated as a sum of duration between the first and second contacts, between the second and third contacts, and between the third and fourth contacts, and is equal to  0.495~days. 

\begin{table*}
\caption{Results for the eclipsing binary stars}              
\begin{tabular}{l c c c}        
\hline\hline                 
ASASSN-V  & J011509.99+652848.0 & J213047.63+161528.6 & J21504 2.53+193829.5 \\
CNAME  &01151017+65284617 &21304769+16152983 &21504261+19382841 \\
$Gaia$ ID&{525129433488850816}&{1772024822529017472}&{1780473641675262848}\\
$\alpha$(2000) [h m s] &01:15:10.17 &21:30:47.69 &21:50:42.61\\
$\delta$(2000) [$^\circ$  $^\prime$  $^{\prime\prime}$]  &+65:28:49.17 &+16:15:29.83 &+19:38:28.41 \\
$Gaia$ $G$ [mag]  &12.10 &12.44 &12.22 \\
Orbital periods [d] &1.31356&2.21       &4.3896       \\
Duration [d] & 0.17 &$0.3\pm0.05$  &$\sim$0.495  \\
Depth [mag] &0.517/0.536&0.882/0.048 &  1.665/0.160 \\
\hline                                   
\end{tabular}
\label{threeEB}
\end{table*}

\subsection{Periodic variable stars}

\begin{figure*}
  \centering
   \includegraphics[width=17 cm]{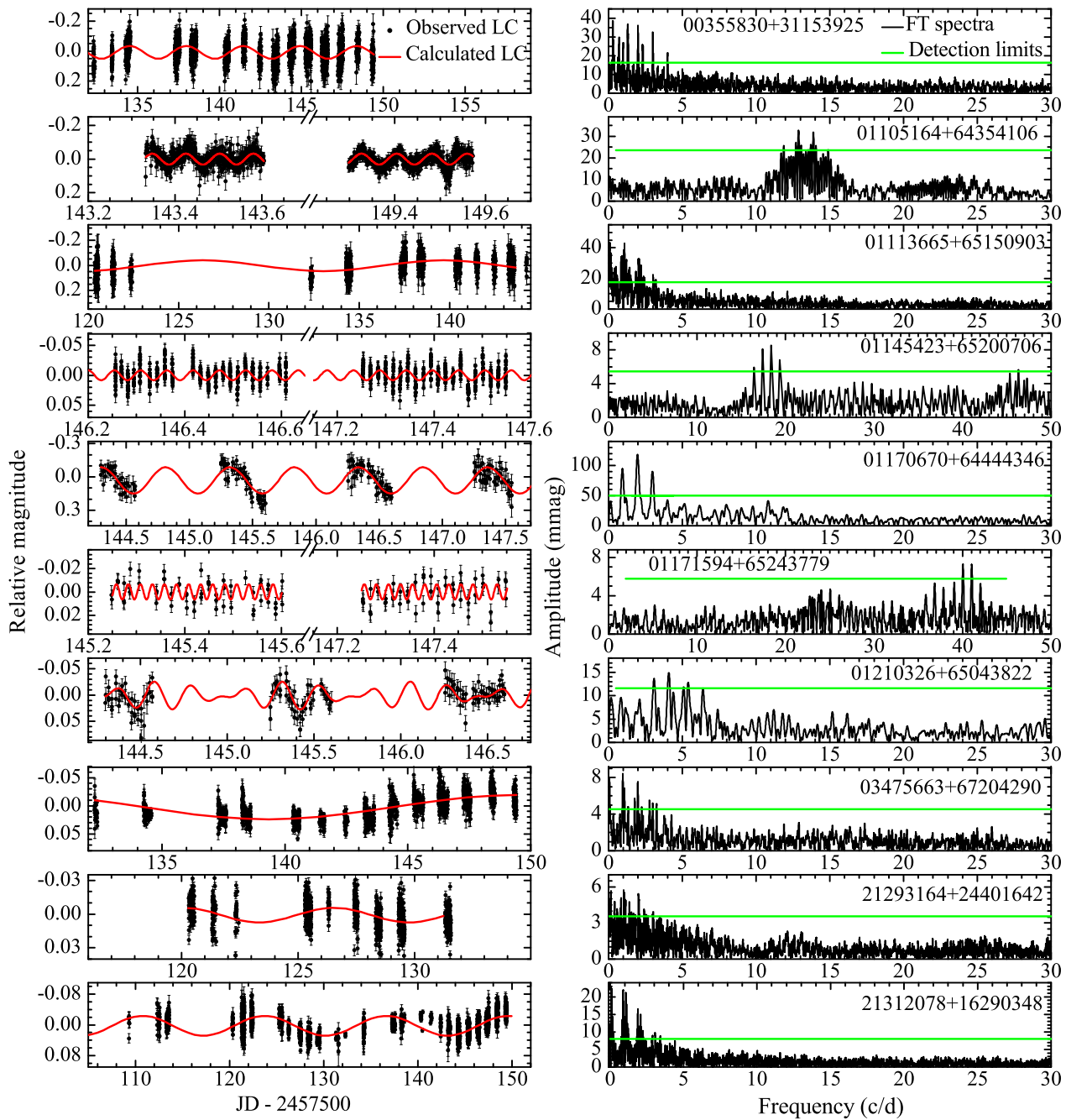}
\caption{Light curves of 10 new-found periodic variable stars (Table~\ref{new-periodic}) and their Fourier amplitude spectra. The red curves correspond to calculated sinusoidal light curves. The green solid lines correspond to the detection limit with FAP=0.01.}
\label{LC_FT_PV}
   \end{figure*}

A group of the observed stars which had a signal in their FT spectra was attributed to periodic variables. We found 10 such stars and derived their oscillation frequencies, amplitudes, and phases using the Period04 code and the method described in Section 3. Fig.~\ref{LC_FT_PV} shows their LCs and FT spectra. 

\begin{table*}
\caption{Results for the new-found periodic variable stars}              
\begin{tabular}{l c c  c c c c}        
\hline\hline                 
  CNAME    & $Gaia$~$G$ &Freq.  &Ampl. 	&Detect.&Phase&Notes		\\
           &    [mag] & [c/d] &  [mmag]	&limit	& & \\	
\hline
00355830+31153925&12.54   &  0.29         &38.99 & 18.16    &0.028& Dwarf\\
01105164+64354106&	11.17&	12.877	&32.37&	23.60 	&0.725&$\delta$~Scuti\\
01113665+65150903&11.90      & 0.08         & 42.83&21.33 &0.607&Red giant\\
01145423+65200706&11.62  & 18.382          & 8.31&5.47 &0.984&$\delta$~Scuti\\
01170670+64444346&	13.55& 	1.96	&119.48&50.30	 	&0.803& EB ? \\
01171594+65243779&11.26   & 40.046          &6.61&5.98  &0.114&$\delta$~Scuti\\
01210326+65043822&	11.81&	4.073	&14.83&	10.60 	&0.561&$\delta$~Scuti\\
	&	 &	5.421	&12.40	& 	&0.198& \\
03475663+67204290&10.00    &0.05      & 21.47 &7.49    &0.045& Dwarf \\
21293164+24401642 &8.42  & 0.16          & 6.44&3.93  & 0.906&  Subgiant or Giant\\
 21312078+16290348&7.67  & 0.08        & 25.93&7.99  & 0.194 &Red giant \\
\hline                                   
\end{tabular}
\label{new-periodic}
\end{table*}   

\begin{figure}[t]
\begin{center}
  \includegraphics[width=\hsize]{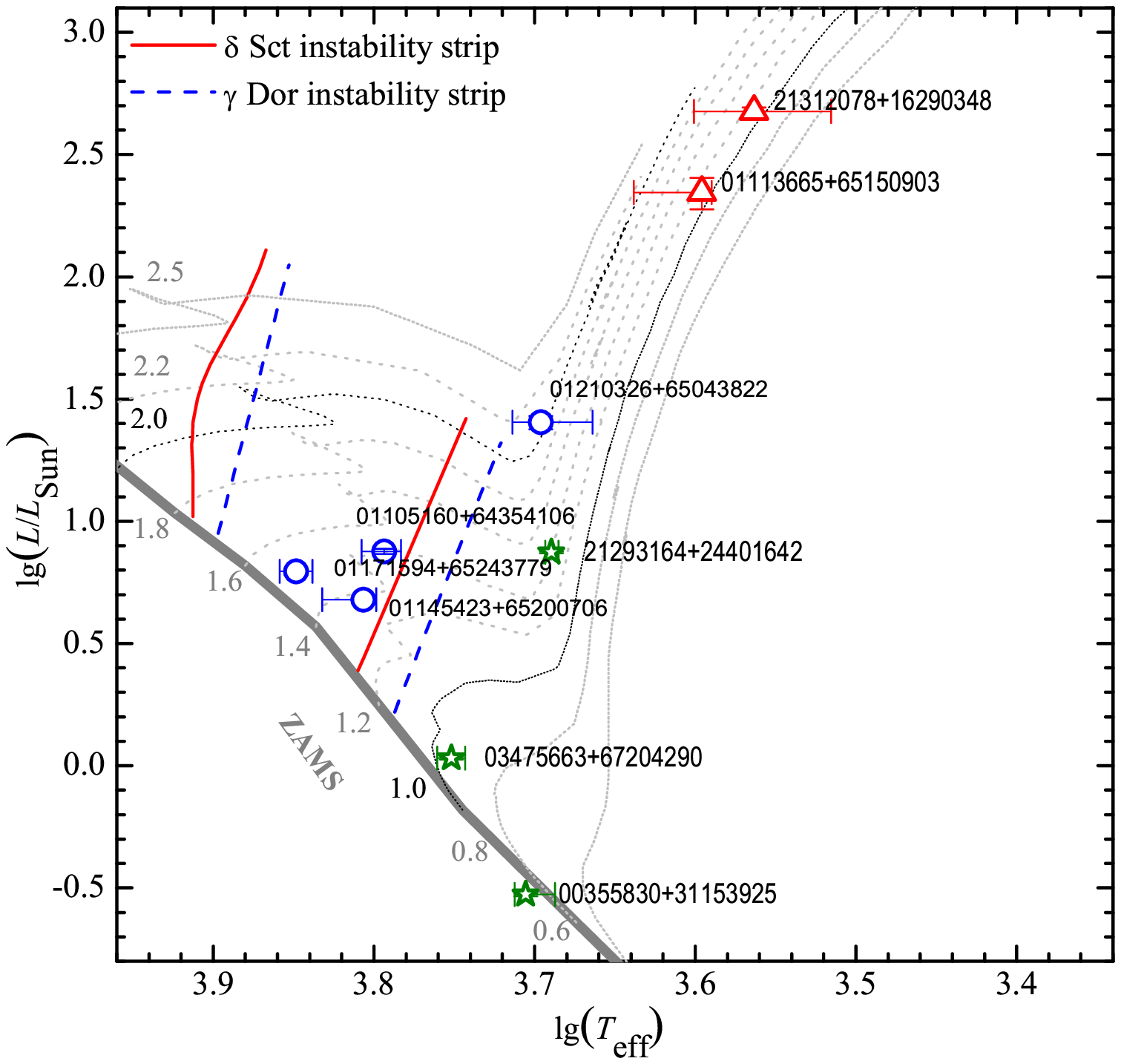}
   \end{center}
\caption{Positions of newly detected variable star candidates in the HR diagram. The open circles correspond to the $\delta$~Scuti candidates, the triangles mark the red giant variables, the stars mark  the variables of indefinite type. Object CNAMEs are presented near the corresponding symbols. The Padova evolutionary tracks with different masses (thin grey lines) and the Zero Age Main Sequence (ZAMS, thick grey line) were taken from {\it http://pleiadi.pd.astro.it/}. 
 Boundaries of instability strips of $\delta$~Scuti and $\gamma$~Doradus type stars were taken from \citet{Xiong2016}. The star 01170670+64444346 is missing in the diagram as no information about its luminosity was found.}
\label{HR_instability}
    \end{figure} 

FT spectra of these stars were calculated using all available light curves. In Table~\ref{new-periodic}, we present frequencies and amplitudes derived from the fitted sinusoidal functions to the observed light curves. 
The highest derived frequency of oscillations is equal to 40.046~c/d, and the lowest one is 0.05~c/d. 

In order to get more information on the new variable stars we picked up their luminosities and effective temperatures from the Gaia~DR2 catalogue where they were determined from parallaxes and three broad-band photometric measurements   (\citealt{GAIA_DR1},  \citealt{DR2_parameters_Andrae}, \citealt{GAIA_DR2_2018}), and plotted  these values with error bars on the Hertzsprung-Russell (HR) diagram (Fig.~\ref{HR_instability}). The star 01170670+64444346 is missing in  Fig.~\ref{HR_instability}, as no information about its luminosity was found in the Gaia~DR2 catalogue, but its lg$T_{\rm{eff}}$ may range from 3.695 to 3.720. This information is not enough to determine the variability type of this star, but its high amplitude (see Table~\ref{new-periodic} and Fig.~\ref{LC_FT_PV}) may indicate that this star is an eclipsing binary with the orbital period shorter than 24 hours.

Four stars showed periodic brightness variations at frequencies which are intrinsic to the $\delta$~Scuti type stars. The lowest frequency of variations at 3.351~c/d belongs to 21511158+19234361, and the highest frequency at 40.046~c/d belongs to 01171594+65243779. The amplitudes vary between 6.61~mmag in 01171594+65243779 and 
119.48~mmag in 01170670+64444346.

Three ~stars~ (01105160+64354106,~ 01145423+ 65200706, and 01171594+65243779) lie inside the $\delta$~Scuti instability strip and are most believably $\delta$~Scuti candidates. 01171594+65243779 probably has the second signal of pulsations at 36.86\,c/d with the amplitude of 5.13\,mmag. As this is slightly below the detection limit, we have not included this frequency into the list of signals. 

Two more stars 
 (01210326+65043822 and 21293164+ 24401642) are outside the instability strip on the red side of the HR diagram. 
For 01210326+65043822 we found more than one frequency above the detection limit. That suggests that this star is a physical variable rather than an eclipsing binary. The 01210326+65043822 is outside the $\delta$~Scuti and $\gamma$~Doradus instability strips at $\sim$2.0~$M_{\odot}$ evolutionary sequence. 
We still classify it as a $\delta$~Scuti candidate, because its two frequencies of brightness variations are intrinsic to such type of stars, and because the $T_{\rm eff}$ value  may be uncertain. 
However, 01210326+65043822 may be also a candidate to RRc Lyrae type stars, which have LCs close to sinusoidal and pulsate with similar frequencies. But according to  \citet{Tsujimoto1998} most RR Lyrae stars are brighter than ${M_{\rm V}}=0.7$~mag or $\lg L/L_{\odot}=1.61$. As luminosity of 01210326+65043822 is  $\lg L/L_{\odot}=1.5$ we classify it as a $\delta$~Scuti candidate too.

All four $\delta$~Scuti candidates should be checked if they are not eclipsing binaries, since the rate of eclipses of some type binaries is close to pulsation frequencies of $\delta$~Scuti stars. Stellar parameters, effective temperatures in particular, should be determined more precisely as well.

The star 21293164+24401642 lies aside from the $\delta$~Scuti and $\gamma$~Doradus instability strips. It may be a subgiant or giant star. It shows low amplitude variability signal at low frequency. 
The rest of the periodic variable candidates were classified according to their position in the HR diagram as dwarfs (00355830+31153925 and 03475663+67204290) or red giants (01113665+65150903 and 21312078+16290348) (see Table~\ref{new-periodic} and Fig.~\ref{LC_FT_PV}). These stars need further observations in order to understand their variability.

 \subsection{Suspected slowly varying stars}
 
We identified 70 stars as possible variable stars with slow changes in brightness or with irregular variability, which cannot be analysed using the Fourier decomposition procedure. We mark these stars in our catalogue as SSVS.
They were selected according to their trending parameters calculated as it was described in Section~3. 

A list of slowly varying stars with their lg(TR$_{\rm O/C})$ parameters is presented in Table~\ref{SVS}, while all other available information on these stars can be found in the online catalogue. 
One of the stars, 01130382+64585393, is a hot B spectral type star with emission lines EM*~AS~12 (\citealt{Skiff2014}, \citealt{Bourges2017}). We found it to be a variable with the period of around 10~days. This may be an early Be type star. 
They are fast rotating stars with the rotation rate close to their critical limit where the centrifugal force balances gravity. All classical Be stars show long-term and gradual variations of the circumstellar
emission and absorption lines, as well as more rapid variability on time scales
ranging from a few minutes to a few days. Short-periodic variations were also found photometrically in some Be stars. In the majority of early-type Be stars, periodic line profile variations can be explained by nonradial pulsations. But there are still a number of Be type stars, which do not allow to make clear conclusions  (\citealt{Porter2003}).

\section{Summary}

\addtolength{\tabcolsep}{12pt}
\begin{table*}
\caption{A list of stars which were selected as slowly varying candidates according to their TR parameter}        
\begin{tabular}{c c c c c}        
\hline\hline                 
CNAME &  lg(TR$_{\rm O/C}$)    &    &CNAME &  lg(TR$_{\rm O/C}$) \\
\hline
01120301+64575428   &   0.110   &&  01204663+64581697	&0.131\\
01130382+64585393   &	0.428	&&  01205769+64490437	&0.792\\
01132435+65215563   &	0.025	&&  01210113+65155781	&0.200\\
01134160+65365858   &	0.243	&&  01210296+64582229	&0.462\\
01141156+64593451   &	0.082	&&  01210501+64582513	&0.167\\
01141948+64554006   &	0.032	&&  01211025+65083869	&0.180\\
01142529+65441374   &	0.028	&&  02212014+41131450	&0.012\\
01142549+65155854   &	0.388	&&  02212557+41053278	&0.136\\
01143371+65220385   &	0.139	&&  02215102+41193841	&0.158\\
01144209+65004113   &	0.031	&&  02215110+41154624	&0.386\\
01150564+64551375   &	0.078	&&  15075542+69545209	&0.204\\
01160384+65183062   &	0.220	&&  15114343+69412625	&0.326\\
01161928+65392227   &	0.009	&&  15120751+69561135	&0.671\\
01163057+64512853   &	0.384	&&  15122671+69451666	&0.015\\
01164025+65084490   &	0.266	&&  20334008+09442108	&0.093\\
01165542+64405716   &	0.367	&&  20340242+10151051	&0.080\\
01172195+65113882   &	0.142	&&  20342002+10160443	&0.316\\
01172629+64555644     & 0.378   &&  21292509+25012481	&0.080\\
01172968+65170489	&0.316	&    &21292844+25025139	&0.069\\ 
01173408+64491585	&0.183	&    &21294677+16163742	&0.041\\ 
01173550+65022366	&0.220	&    &21294758+24360862	&0.051\\ 
01180348+65123308	&0.114	&    &21295664+24361997	&0.253\\ 
01183184+65011706	&0.017  &    &21300347+24391585	&0.055\\ 
01184581+65214439	&0.063	 &   &21310240+16180271 	&0.294\\
01185356+65162016	&0.164	   & &21311019+24280300	&0.187\\
01190383+65135887	&0.117	   & &22580219+34213263	&0.232\\
01191419+65003688	&0.222	   & &22584679+34222375	&0.278\\
01192642+64431285	&0.323	   & &23172777+36085709	&0.186\\
01192860+64553329	&0.270	   & &23173360+36001939	&0.102\\
01195377+65102501	&0.141	   & &23173742+35541606	&0.123\\
01200267+64494614	&0.012	 &   &23175486+35473659	&0.109\\
01201787+64454215	&0.124	 &   &23183110+35585173	&0.029\\
01202094+64432590	&0.039	 &   &23192371+36200998	&0.040\\
01202623+65141597	&0.041	 &   &23200080+36023470	&0.266\\
01204136+64510537	&0.338	 &   &23291730+19331353	&0.146\\ 
\hline                                   
\end{tabular}
\label{SVS}
\end{table*}
\addtolength{\tabcolsep}{-12pt}    

Since 2016, when we started our Spectroscopic and Photometric Survey of the Northern Sky (SPFOT; Mikolaitis et al. 2018) at the Mol\.{e}tai Astronomical Observatory, we obtained 24\,470 CCD images and analysed stellar light curves of 3598 stars in 13 fields of the northern sky (in total 5.85\,deg$^2$).  

We found 81 new variable stars (one eclipsing binary, four $\delta$~Scuti candidates, six other variables with periods between 35 minutes and 20 days, and 70 slowly varying stars with so far undefined periodicity). Additional photometric and spectral observations were carried out for  TYC\,2764-1997-1, 
 previously considered as a candidate for eclipsing contact binaries, and its status confirmed. Complementary observations were performed for the eclipsing binaries ASASSN-V J011509.99+652848.0, J213047.63+161528.6, and J215042.53+193829.5       previously listed in the ASAS-SN data base (\citealt{Jayasinghe2018_II}, \citealt{Jayasinghe2018}). 
 All the calibrated images, obtained light curves and determined parameters are presented in an online catalogue.

Observations of these objects will have to be continued, with a special focus on spectroscopic spectral type determinations. Radial velocity measurements and multi-colour photometric observations are also of high importance for modelling the eclipsing binary systems and for recognising close binary systems.

\acknowledgments{This research was funded by the Research Council of Lithuania (LAT-08/2016). We thank the anonymous referee for comments and
suggestions, which helped to improve this Paper. We acknowledge observing time with the Nordic Optical Telescope, operated by the Nordic Optical Telescope Scientific Association at the Observatorio del Roque de los Muchachos, La Palma, Spain, of the Instituto de Astrofisica de Canarias. We acknowledge observations with ALFOSC, which is provided by the Instituto de Astrofisica de Andalucia (IAA) under a joint agreement with the University of Copenhagen and NOTSA. This research has made use of the SIMBAD database and NASA's Astrophysics Data System (operated at CDS, Strasbourg, France).}

\bibliographystyle{spr-mp-nameyear-cnd.sty}
\bibliography{biblio-u1}

\end{document}